\documentclass[preprint,showpacs,preprintnumbers,amsmath,amssymb]{revtex4}
\usepackage{graphicx}
\newcommand{\al}{\alpha}
\newcommand{\be}{\beta}
\newcommand{\pa}{\partial}

\newcommand{\si}{\sigma}

\newcommand{\de}{\delta}

\newcommand{\De}{\Delta}

\newcommand{\tha}{\theta}

\newcommand{\rar}{\rightarrow}

\newcommand{\non}{\nonumber}
\begin{document}

\preprint{Preprint LPT-ORSAY 01-67 /  M\'exico ICN-UNAM 01-15}

\title{H$_2^+$ ion in a strong magnetic field:\\
 Lowest gerade and ungerade electronic states }

\author{A.~V.~Turbiner}
 \altaffiliation[]{On leave of absence from the Institute for Theoretical
 and Experimental Physics, Moscow 117259, Russia.}
\email{turbiner@lyre.th.u-psud.fr}
\affiliation{%
 Laboratoire de Physique Theorique, B${\hat a}$t.210, Universit\'e
 Paris-Sud, Orsay, F-91405 France \\ and \\
Instituto de Ciencias Nucleares, Universidad Nacional
Aut\'onoma de M\'exico, Apartado Postal 70-543, 04510 M\'exico, D.F., Mexico.}
\author{J.~C.~L\'opez V.}
\email{vieyra@nuclecu.unam.mx}
\affiliation{%
Instituto de Ciencias Nucleares, Universidad Nacional
Aut\'onoma de M\'exico, Apartado Postal 70-543, 04510 M\'exico, D.F.,
Mexico.}

\author{A.~Flores-Riveros}
\email{flores@sirio.ifuap.buap.mx}
\affiliation{%
Instituto de F\'{\i}sica, Benem\'erita Universidad
Aut\'onoma de Puebla, Apartado Postal J-48, 72570 Puebla, Pue.,
Mexico.}

\date{November 8, 2001}

\begin{abstract}
  In the framework of a variational method with a {\it single} trial
  function an accurate study of the lowest gerade $1_g$ and ungerade
  $1_u$ electronic states of the molecular ion $H_2^+$ in a magnetic
  field is performed. Magnetic field ranges from $0$ to $4.414 \times
  10^{13}$ G and orientations of the molecular axis with respect to
  the magnetic line $0^{\circ} \leq \tha \leq 90^{\circ}$ are
  considered.  A one-parameter gauge dependent vector potential is
  used in the Hamiltonian, which is finally variationally optimized.
  A well pronounced minimum on the total energy surface of the $(ppe)$
  system in both $1_g$ and $1_u$ states is found for all magnetic
  fields and orientations studied. It is shown that for both states
  the parallel configuration ($\tha=0$) at equilibrium always
  corresponds to the minimal total energy. It is found that for a
  given $\tha$ for both states the magnetic field growth is always
  accompanied by an increase in the total and binding energies as well
  as a shrinking of the equilibrium distance. We demonstrate that for
  $B\gtrsim 1.8 \times 10^{11}$ G the molecular ion can dissociate,
  $H_2^+ \rar H\ +\ p$, over a certain range of orientations
  ($\tha_{cr} \le \tha \le 90^{\circ}$), where the minimal
  $\tha_{cr}\simeq 25^{\circ}$ occurs for the strongest magnetic field
  studied, $B=4.414 \times 10^{13}$ G.  For $B<10^{12}$ G the ion
  $H_2^+$ in $1_g, 1_u$ states is the most compact, being in the
  perpendicular configuration ($\tha$=90$^{\circ}$), whereas for
  $B\gtrsim 10^{12}$ this occurs for an angle $<90^{\circ}$.  For the
  $1_g$ state in any orientation, with the magnetic field growth at $B
  \sim 10^{11}$ G, a two-peak electronic distribution changes to
  single-peak one.
\end{abstract}


\maketitle

\section{Introduction}

There are theoretical qualitative indications that in the presence of
a strong magnetic field the physics of atoms and molecules gets a
wealth of new, unexpected phenomena even for the simplest systems
\cite{Kadomtsev:1971,Ruderman:1971}. In practice, the atmosphere of
neutron stars, which is characterized by the existence of enormous
magnetic fields $10^{12} - 10^{13}$ G, provides a valuable paradigm
where this physics could be realized.

One of the first general features observed in standard atomic and
molecular systems placed in a strong magnetic field is an increase in
the binding energy, accompanied by a drastic shrinking of the electron
localization length. It leads to a decrease of the equilibrium
distance with magnetic field growth. This behavior can be considered
as a consequence of the fact that at large magnetic fields the
electron cloud takes a needle-like form extended along the magnetic
field direction and the system becomes effectively
quasi-one-dimensional.  Furthermore the phenomenon of
quasi-one-dimensionality enhances the stability of standard atomic and
molecular systems. They become elongated along the magnetic line
forming molecules of the type of  linear polymers (for details see review
papers \cite{Garstang:1977, Lai:2001}). It also hints at the existence
of exotic atomic and molecular systems which do not exist in the
absence of a magnetic field.  Motivated by these simple observations
it was shown in \cite{Turbiner:1999, Lopez-Tur:2000} that exotic
one-electron molecular systems $H_3^{++}$ and $H_4^{+++}$ can exist in
sufficiently strong magnetic fields in the form of linear polymers.
However, the situation becomes much less clear (and also much less
investigated) when the nuclei are not aligned with the magnetic field
direction, thus not forming in general a linear system. Obviously,
such a study would be important for understanding the kinetics of a
gas of molecules in the presence of a strong magnetic field. As a
first step towards such a study, even the simplest molecules in
different spatial configurations deserve attention. The goal of the
present work is to attempt to make an extensive quantitative
investigation of the lowest electronic states of H$_2^+$ in a
(near)equilibrium position in the framework of a single approach in
its entire complexity: wide range of magnetic field strengths ($0 - 4
\times 10^{13}$ G) and arbitrary orientation of the molecular axis
with respect to magnetic line.

It is well known that the molecular ion H$_2^+$ is the most stable
one-electron molecular system in the absence of a magnetic field. It
remains so in the presence of a constant magnetic field as well, but
it turns out that for a magnetic field $B > 10^{13}$ G the exotic ion
H$_3^{++}$ appears to be the most bound (see \cite{Lopez-Tur:2000}).
The ion H$_2^+$ has been widely studied, both with and without the
presence of a magnetic field, due to its importance in astrophysics,
atomic and molecular physics, solid state and plasma physics (see
[\onlinecite{Garstang:1977}-\onlinecite{Turbiner:2001}] and references
therein). The majority of the previously performed studies was focused
on the case of the parallel configuration, when the angle between the
molecular axis and the magnetic field direction was zero, $\tha=0$,
with an exception of \cite{Schmelcher}, where a detailed quantitative
analysis was performed for any $\tha$ but for $B=1a.u.$ Previous
studies were based on various numerical techniques, most of all --
different versions of the variational method, including the
Thomas-Fermi approach. As a rule, in these studies the nuclear motion
was separated from the electronic motion using the Born-Oppenheimer
approximation at zero order -- assuming protons as infinitely heavy
charged centers. It was observed at a quantitative level that magnetic
field growth is always accompanied by an increase in the total and
binding energies, as well as shrinking of the equilibrium distance.
As a consequence it led to a striking conclusion about the drastic
increase in the probability of nuclear fusion for H$_2^+$ in the
presence of a strong magnetic field \cite{Kher}.

In the present study we perform accurate calculations for the lowest
$1_g$ and $1_u$ electronic states of H$_2^+$ in (near)equilibrium
position for magnetic fields $B = 0 - 4.414 \times 10^{13}$ G and
arbitrary orientation of the molecular axis towards the magnetic line.
Since the Hamiltonian is gauge-dependent a choice of the form of the
vector potential is one of the crucial points in our study. We
construct state-of-the-art, `adequate', trial functions with a
variationally optimized gauge dependence consistent with the choice of
vector potential. Although an appropriate position of the gauge
origin\footnote{where the vector potential vanishes, see Section 2}
may be important (especially for large internuclear distances, this
lies beyond the scope of the present article and will be discussed
elsewhere) we place the gauge origin in the middle between the two
nuclei (charged centers) and keep it fixed. For the parallel
configuration the present work can be considered as an extension (and
also an improvement) of previous work \cite{Lopez:1997}.  It is
necessary to emphasize that we encountered several new physical
phenomena which occur when the molecular axis deviates from the
magnetic field direction. In particular, if the magnetic field is
sufficiently strong, $B\simeq 10^{11}$G the ion H$_2^+$ can dissociate
to $H + p$ \footnote{It confirms and extends a prediction made by
  Larsen \cite{Larsen} about the possible instability of H$_2^+$ in
  the perpendicular configuration for magnetic fields $B \gtrsim 1.6
  \times 10^{11}$ G. The same phenomenon was mentioned by Khersonskij
  \cite{Kher} for B$\gtrsim$10$^{12}$ G. An accurate study of this
  phenomenon was carried out in Ref.\cite{Turbiner:2001}.}.  This
means that even though the positive binding energy of H$_2^+$ in the
optimal configuration ensures its existence, even under the high
temperature conditions prevailing on the surface of neutron stars
$(10-100\ eV)$, there is a certain probability of dissociation.  The
behavior of the equilibrium distance as a function of $\tha$ reveals
another surprising feature: for magnetic fields B$<10^{12}$ G the
molecule is most compact in the perpendicular configuration, while for
$B\gtrsim 10^{12}$ G this occurs for a certain angle $\tha <90^o$.  We
find that for the $1_g$ state at any orientation in the weak field
regime the electronic distribution peaks at the positions of the
protons, while at large magnetic fields the electronic distribution is
characterized by one maximum which occurs at the midpoint between two
protons. This change appears around $B \sim 10^{10-11}$ G with a
slight dependence on the inclination angle $\tha$. From a physical
point of view the former means that the electron prefers to stay in
the vicinity of a proton. It can be interpreted as a dominance of the
$H$-atom plus proton interaction.  The latter situation implies that
the electron is `shared' by both protons and hence a separation to
$H$-atom plus proton cannot be done.  Therefore, we can call the
two-peak situation an ionic coupling, while the one-peak case is
assigned to the covalent coupling, although this definition differs
from the one widely accepted in textbooks (see for example \cite{LL}).
Thus, we can conclude that a new phenomenon appears - as the magnetic
field grows the type of coupling changes from ionic to covalent.

Atomic units are used throughout ($\hbar$=$m_e$=$e$=1) albeit
energies are expressed in Rydbergs (Ry). In particular, the
magnetic field $B$ is given in a.u. with $B_0= 2.35 \times 10^9 G$.

\section{Theory}

The Hamiltonian which describes the H$_2^+$ molecular ion placed
in a uniform constant magnetic field directed along the $z$-axis,
${\cal B}=(0,0,B)$ is given by (see, for example, \cite{LL})
\begin{equation}
\label{Ham}
 {\cal H} = {\hat p}^2 + \frac{2}{ R} -\frac{2}{r_1} -\frac{2}{r_2}  -
({\hat p} {\cal A}) +  {\cal A}^2 \ ,
\end{equation}
(see Fig.1 for notations), where ${\hat p}=-i \nabla$ is the
electron momentum, ${\cal A}$ is a vector potential corresponding
to the magnetic field $\cal B$.

\begin{figure}
\includegraphics*[width=2.0in,angle=-90]{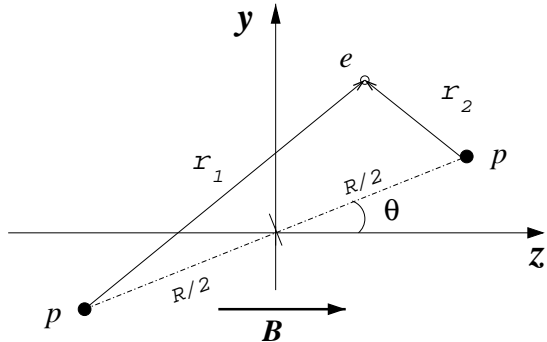}%
\caption{\label{fig:1} Geometrical setting  for the H$_2^+$ ion in a
magnetic field directed along the $z$-axis.}
\end{figure}

The vector potential is gauge-dependent and hence it is defined
ambiguously. Therefore the Hamiltonian (\ref{Ham}) is also explicitly
gauge-dependent although the energies and other observables do not.
Thus, when solving the eigenvalue problem corresponding to the above
Hamiltonian within an approximate method, the energies (as well as
other observables) will in general be gauge-dependent (only the exact
ones would be gauge-independent).  Hence, one can choose the form of
the vector potential in a certain optimal way, for instance, which
would lead to minimal approximate total energy of the ground state. In
order to realize this idea let us take a certain one-parametric family
of vector potentials corresponding to the constant magnetic field $B$
(e.g. see \cite{Larsen})
\begin{equation}
\label{Vec}
  {\cal A}= B((\xi-1)y, \xi x, 0)\ ,
\end{equation}
where $\xi$ is the parameter to be chosen in an optimal way.
The well known symmetric (or circular)  and asymmetric gauges
are particular cases of (\ref{Vec}) corresponding to $\xi$=1/2 and
$\xi=1$, respectively (see, for instance, \cite{LL}). By
substituting (\ref{Vec}) into (\ref{Ham}) we arrive at the
Hamiltonian
\begin{widetext}
\begin{equation}
\label{Ham.fin}
 {\cal H} = -{\nabla}^2 + \frac{2}{ R} -\frac{2}{r_1} -\frac{2}{r_2}  +
i B[(\xi-1)y\pa_x + \xi x\pa_y] +  B^2 [ \xi^2 x^2+ (1-\xi)^2
y^2] \ .
\end{equation}
\end{widetext}

The idea of choosing an optimal gauge is widely exploited in quantum
field theory. Usually, the gauge is fixed following a certain
convenience criterion or for technical simplicity.  Regarding our
present problem of H$_2^+$ in a magnetic field, there were also 
some attempts to discuss the gauge fixing (see, for instance,
\cite{Schmelcher} and references therein). Perhaps the first
constructive (and remarkable) attempt was realized by Larsen
\cite{Larsen} in his study of the ground state. In particular, he
explicitly showed that gauge dependence of the variational results can be
quite significant and even an oversimplified optimization procedure
improves drastically the accuracy of the numerical results.

It is rather evident that there exists a certain gauge for which the
Hamiltonian (\ref{Ham}) has the ground state eigenfunction given by a
real function \footnote{ Similar statement is correct for any other
  given eigenfunction. In general, reality of the eigenfunctions
  corresponds to different values of gauge parameter}.  Thus, we can
use real trial functions with explicit dependence on the gauge
parameter $\xi$. {\em This parameter is fixed by performing a
  variational optimization of the energy}. Therefore, as a result of
the minimization we find both a variational energy and a gauge for
which the eigenfunction is real, as well as the corresponding
Hamiltonian. One can show that for a system possessing axial
(rotational) symmetry \footnote{ This is the case whenever the
  magnetic line is directed alongside the molecular axis (parallel
  configuration)} the optimal gauge is the symmetric one $\xi=1/2$. It
is precisely this gauge which was overwhelmingly used (without any
explanations) in the majority of previously performed H$_2^+$ studies
in the parallel configuration [3-22,25].  However, this is not the case
if $\tha \neq 0^{\circ}$.  For the symmetric gauge the exact
eigenfunction now becomes complex, hence complex trial functions
should be used. But we can avoid complexity by adjusting the gauge so
as to continue with real eigenfunctions. This justifies the use of
real trial functions.  Our results (see Section 3) lead to the
conclusion that for both studied states the optimal gauge parameter is
$\xi \in [0.5,1]$.  In the limit $B \gg 1a.u.$, the parameter $\xi$
tends to one corresponding to asymmetric gauge. Likely, this tendency
will continue for other excited states.

One can easily see that the expectation value of the term $\sim B$ in
Eq.(\ref{Ham.fin}) vanishes when it is calculated with respect to any
{\it real} normalizable function. Hence, without loss of generality,
this term can be omitted in the Hamiltonian (\ref{Ham}). It gives an
essential simplification of the computational procedure. Our choice of
variational trial function $\Psi_0$ can be formulated as follows
\cite{Tur}: {\sl we construct a real trial function for which the
  potential $V_0=\frac{\De \Psi_0}{\Psi_0}$ reproduces the original
  potential near Coulomb singularities and its growing terms at large
  distances; it also supports all symmetries of the problem at hand
  and the gauge parameter $\xi$ is included into the trial function
  explicitly.}  This recipe (for symmetric gauge) was successfully
exploited in the previous study of H$_2^+$ in strong magnetic fields
in the parallel configuration \cite{Lopez:1997}. This recipe was also
used to construct trial functions when the question of the existence
of the exotic ions H$_3^{++}$ and H$_4^{+++}$ is considered
\cite{Turbiner:1999,Lopez-Tur:2000}.

The Hamiltonian (\ref{Ham}) gives rise to different symmetry
properties depending on the orientation of the magnetic field with
respect to the internuclear axis. The most symmetric situation
corresponds to $\tha=0^{\circ}$, where invariance under permutation of
the (identical) charged centers $P: (1\leftrightarrow 2)$ together
with $P_z: (z \to -z)$ symmetry hold. Since the angular momentum
projection $\ell_z =m$ is conserved, $P_z$ accounts also for the
degeneracy $m\to -m$. Thus, we classify the states as $ 1\si_{g,u},
2\si_{g,u}, \ldots 1\pi_{g,u}, 2\pi_{g,u} \ldots 1\de_{g,u},
2\de_{g,u} \ldots$, where the numbers $1,2,\ldots$ refer to the
electronic states in increasing order of energy. The labels
$\si,\pi,\de \ldots$ are used to denote $|m|=0,1,2 \ldots$,
respectively, the label $g$ ($u$) gerade (ungerade) is assigned to the
states of even (odd) parity $P$ of the system. At $\tha=90^o$ the
Hamiltonian still remains invariant under the parity operations $P$
and $P_z$, while the angular momentum projection is no longer
conserved and $m$ is no longer a quantum number. The classification in
this case is $1_{g,u}^\pm, 2_{g,u}^\pm, \ldots$, where sign $+(-)$ is
reserved to denote even (odd) $z$-parity.  Eventually, for arbitrary
orientation the only parity under permutations $P$ is conserved.  In
general, we refer to the lowest gerade and ungerade states in our
study as $1_g$ and $1_u$.  This is the only notation which make sense
for all orientations $0^{\circ} \leq \tha \leq 90^{\circ}$.

One of the simplest trial functions for $1_g$ state which meets the
requirements of our criterion of adequacy is
\begin{eqnarray}
\label{tr:1}
 \Psi_1= {e}^{-\al_1  (r_1+r_2)}
 {e}^{- B  [\be_{1x} \xi x^2 + \be_{1y}(1-\xi) y^2] }\,
\end{eqnarray}
(cf. \cite{Lopez:1997, Turbiner:2001}), where $\al_1$, $\be_{1x}$ and
$\be_{1y}$ are variational parameters and $\xi$ is the parameter of
the gauge (2).  The first factor in this function, being symmetric
under interchange of the charged centers $r_1\leftrightarrow r_2$, and
corresponding to the product of two $1s$-Coulomb orbitals centered on
each proton, is nothing but the celebrated Heitler-London
approximation for the ground state $1\si_g$. The second one is the
lowest Landau orbital corresponding to the vector potential of the
form Eq. (\ref{Vec}). So, the function (\ref{tr:1}) can be considered
as a modification of the free field Heitler-London function. Following
the experience gained in studies of H$_2^+$ without magnetic field it
is natural to assume that Eq. (\ref{tr:1}) is adequate to describe
internuclear distances near equilibrium. This assumption will be
checked (and eventually confirmed) {\it a posteriori}, after making
concrete calculations (see Section 3).

The function (\ref{tr:1}) corresponds to an exact eigenfunction of
the potential
\begin{widetext}
\begin{eqnarray}
  \label{V:1}
  V_1 &=& \frac{\nabla^2 \Psi_1}{\Psi_1} =
2 \al_1^2 -\frac{B}{2}[\be_{1x}\xi+\be_{1y}(1-\xi)] +
\frac{B^2}{4} [ \be_{1x}^2 \xi^2 x^2 + \be_{1y}^2(1-\xi)^2 y^2 ] +
2 \al_1^2 ({\hat n_1}\cdot{\hat n_2}) \non \\
 && + {\al_1} B \left[
\frac{\be_{1x}\xi x (x-x_1) + \be_{1y}(1-\xi) y (y-y_1)}{r_1} +
\frac{\be_{1x}\xi x (x-x_2) + \be_{1y}(1-\xi) y (y-y_2)}{r_2}
\right] 
\nonumber \\ &&
-2 \al_1 \left[ \frac{1}{r_1} + \frac{1}{r_2}
\right]. \non
\end{eqnarray}

The potential $V_1$ reproduces the functional behaviour of the
original potential (\ref{Ham.fin}) near Coulombic singularities
and at large distances. These singularities are reproduced exactly
when $\be_{1x}=\be_{1y} =1/2$ and $\al_1=1$.
\end{widetext}
Another trial function meets the requirements of our
criterion of adequacy as well  
\begin{equation}
\label{tr:2}
\Psi_2= \bigg({e}^{-\al_2 r_1} + P {e}^{-\al_2 r_2}\bigg)
{e}^{ -  B  [\be_{2x}\xi x^2 +\be_{2y}(1-\xi) y^2] }\ ,
\end{equation}
(cf. \cite{Lopez:1997, Turbiner:2001}).  It is the celebrated
Hund-Mulliken function of the free field case multiplied by the lowest
Landau orbital, where $\al_2$, $\be_{2x}$ and $\be_{2y}$ are
variational parameters. The parameter $P$ has the meaning of the
parity, being equal to $+1$ for $1_g$ and $-1$ for $1_u$. From a
physical point of view this function describes the interaction between
a hydrogen atom and a proton (charged center), and also models the
possible decay mode of H$_2^+$ into hydrogen atom plus proton.  Thus,
one can naturally expect that for sufficiently large internuclear
distances $R$ this function prevails giving a dominant contribution.
Again this assumption will be checked {\it a posteriori}, by concrete
calculations (see Section 3).

There are two natural ways to incorporate the behavior of the system
both near equilibrium and at large distances into a single trial
function. It is to make a linear or nonlinear interpolation. The linear
interpolation is given by a linear superposition
\begin{equation}
\label{tr:3a}
\Psi_{3a}= A_1 \Psi_{1} + A_2 \Psi_{2}\ ,
\end{equation}
where $A_1$ or $A_2$ are parameters and one of them is kept fixed by
a normalization condition. The function (\ref{tr:3a}) makes sense
for $1_g$ state only since the function (\ref{tr:1}) does not exist
for $1_u$ state. Therefore, conditionally, for state $1_u$ we simply
put the parameter $A_1=0$. In turn, the simplest nonlinear
interpolation is of the form
\begin{equation}
\label{tr:3b}
\Psi_{3b}= \bigg({e}^{-\al_3 r_1-\al_4 r_2} +
P {e}^{-\al_3 r_2-\al_4 r_1}\bigg)
{e}^{ - B  [\be_{3x}\xi x^2 +\be_{3y}(1-\xi) y^2] }\ ,
\end{equation}
(cf. \cite{Lopez:1997, Turbiner:2001}), where $\al_{3}$, $\al_{4}$,
$\be_{3x}$ and $\be_{3y}$ are variational parameters and $P=\pm 1$ is
the parity of the state. This is a Guillemin-Zener function of the
free field case multiplied by the lowest Landau orbital. If
$\al_3=\al_4$, the function (\ref{tr:3b}) coincides with (\ref{tr:1}).
If $\al_4=0$, the function (\ref{tr:3b}) coincides with (\ref{tr:2}).

The most general Ansatz is a linear superposition of trial
functions (\ref{tr:3a}) and (\ref{tr:3b}),
\begin{equation}
\label{trial}
\Psi = A_1 \Psi_{1} + A_2 \Psi_{2}+A_{3}\Psi_{3b}\ ,
\end{equation}
where we fix one of the $A$'s and let all other parameters vary.  For
the $1_u$ state we put the parameter $A_1=0$ since the function
(\ref{tr:1}) does not exist for this state. Finally, the total number
of variational parameters in (\ref{trial}), including $R$ and $\xi$,
is fourteen for the $1_g$ state and ten for the $1_u$ state,
respectively. For the parallel configuration, $\tha=0^{\circ}$, the
parameter $\xi=0.5$ and hence $\beta_{1x}=\beta_{1y},
\beta_{2x}=\beta_{2y}, \beta_{3x}=\beta_{3y}$, and the number of free
parameters is reduced to ten for the $1_g$ state and seven for the
$1_u$ state, respectively.  Finally, with the function (\ref{trial})
we intend to describe the lowest (un)gerade state for {\it all}
magnetic fields where non-relativistic consideration is valid for $B
\leq 4.414 \times 10^{13}$ G, and for {\it all} orientations of the
molecular axes.

Our variational calculations were performed by using the
minimization package MINUIT of CERN-LIB. Numerical integrations
were carried out with relative accuracy $\sim 10^{-7}$ by use of 
adaptive NAG-LIB  (D01FCF) routine. All calculations were
performed on a Pentium-III PC $750MHz$.

\section{Results and discussion}

In this Section we present the results for
the $1_g$ and the $1_u$ states of H$_2^+$ for magnetic fields
ranging from $B=0$ G through $B=4.414 \times 10^{13}$ G, where a
non-relativistic consideration is relevant and relativistic
corrections can be neglected (see \cite{Salpeter:1996} for
discussion), and for orientations ranging from $0^{\circ}$ (parallel
configuration) to $90^{\circ}$ (perpendicular configuration).

\subsection{$1_g$ state}

Before coming to a concrete quantitative consideration we must state
that for all explored magnetic fields, $B = 0 - 4.414 \times 10^{13}$
G, and all orientations we found that a well pronounced minimum in the
total energy of the system $(ppe)$ appears at {\it finite}
internuclear distance. This minimum always corresponds to positive
binding energy $E_b$ \footnote{The binding energy is defined as the
  affinity to keep the electron bound, $E_b = B-E_T$.  $B$ is given in
  $Ry$ and thus has the meaning of the energy of a free electron in a
  magnetic field.}  and hence it implies the {\it existence} of the
H$_2^+$ molecular ion for magnetic fields $B = 0 - 4.414 \times
10^{13}$ G and any orientation of the molecular axis.  This is at variance
with the statement by Khersonskij \cite{Kher} about the {\it non-existence}
of a minimum at finite distances on the total energy surfaces at
sufficiently strong magnetic fields for some, far from parallel
orientations and therefore {\it non-existence} of the molecular ion
H$_2^+$.  Presumably, this statement of non-existence is an artifact
arising from an inappropriate choice of trial functions and therefore
a consequent loss of accuracy.  It is worth emphasizing that the
variational study in \cite{Kher} was carried out with a trial function
somewhat similar to that of Eq.(\ref{tr:2}), but which does not
fulfill our criterion of adequacy in full. The potential corresponding
to this function reproduces correctly the original potential near
Coulomb singularities and $\sim \rho^2$-growth at large distances.
However, it generates growing terms $\sim \rho$ which implies a
reduction in the rate of convergence of a perturbation theory for which
the variational energy represents the first two terms (see discussion in
\cite{Tur}).

In Tables I, II and III the results for the total energy $(E_T)$,
binding energy $(E_b)$ and equilibrium distance $(R_{eq})$ are shown
for $\tha$=0$^{\circ}$, 45$^{\circ}$ and 90$^{\circ}$, respectively.
As seen in Table I, our results for $\tha=0^{\circ}$ lead to the
largest binding energies for $B > 10^{11}$ in comparison with other
calculations. As for $B \lesssim 10^{11}$ G, our binding energies for
the parallel configuration appear to be very close (of the order of
$\lesssim 10^{-4-5}$ in relative deviation) to the variational results
at Wille \cite{Wille:1988}, which are the most accurate so far in this
region of magnetic field strengths.  They are based on a trial
function in the form of a linear superposition of $\lesssim 500$
Hylleraas type functions.  It is quite amazing that our simple trial
function (8) with ten variational parameters gives comparable (for $B
\lesssim 10^{11}$ G) or even better (for $B > 10^{11}$ G) accuracy.
It is important to state the reason why the trial function
\cite{Wille:1988} fails being, increasingly inaccurate as a function
of magnetic field growth for $B>10^{11}$ G. An explanation of this
inaccuracy is related to the fact that in the $(x,y)$- directions the
exact wave function decays asymptotically as a Gaussian function, unlike
Hylleraas functions which decay like the exponential of a linear function.
The potential corresponding to the function \cite{Wille:1988}
reproduces correctly the original potential near Coulomb singularities
but fails to reproduce $\sim \rho^2$-growth at large distances. It
implies a zero radius of convergence of the perturbation theory for
which the variational energy represents the first two terms (see
discussion in \cite{Tur}).

\begingroup
\squeezetable
\begin{table*}
\caption{\label{Table:1}
   Total $E_T$, binding $E_b$ energies and 
   equilibrium distance $R_{eq}$  for the state $1_g$ in parallel
  configuration,  $\tha=0^{\circ}$. ${}^{\dagger}$~This value is taken from 
 \cite{Lopez:1997}}
\begin{ruledtabular}
\begin{tabular}{lcccl}
    $B$ &  $E_T$ (Ry) &  $E_{b}$ (Ry) & $R_{eq}$ (a.u.) &  \\
    \hline
  $B=0$
               & -1.20525  & ---    & 1.9971 &  Present${}^{\dagger}$\\
               & -1.20527  & ---    & 1.997  &  Wille \cite{Wille:1988}\\
  $ 10^{9} $G
               & -1.15070  & 1.57623    & 1.924 &  Present\\
               & -1.15072  & 1.57625    & 1.924  &  Wille \cite{Wille:1988}\\
    $ 1$ a.u.  & -0.94991  & 1.94991    & 1.752  &  Present\\
               & ---       & 1.9498     & 1.752  &  Larsen \cite{Larsen}\\
               & -0.94642  & 1.94642    & 1.76   &  Kappes et al \cite{Schmelcher}\\
 $ 10^{10} $ G
               & 1.09044   &  3.16488   & 1.246  &  Present\\
               & 1.09031   &  3.16502   & 1.246  &  Wille \cite{Wille:1988}\\
    $ 10$ a.u  & 5.65024   &  4.34976   & 0.957  & Present\\
               & ---       &  4.35      & 0.950  & Wille \cite{Wille:1988}\\
               & ---       &  4.35      & 0.958  & Larsen \cite{Larsen}\\
               & ---       &  4.3346    & 0.950  & Vincke et al \cite{Vincke}\\
  $ 10^{11}$ G
               & 35.0434   & 7.50975    & 0.593  &  Present\\
               & 35.0428   & 7.5104     & 0.593  &  Wille \cite{Wille:1988}\\
               & ---       & 7.34559    & 0.61   &  Lai et al \cite{Salpeter:1992} \\
 $ 100 $ a.u.  & 89.7096   & 10.2904    & 0.448  & Present\\
               & ---       & 10.2892    & 0.446  & Wille \cite{Wille:1988}\\
               & ---       & 10.1577    & 0.455  & Wunner et al \cite{Wunner:82}\\
               & ---       & 10.270     & 0.448  & Larsen \cite{Larsen}\\
               & ---       & 10.2778    & 0.446  & Vincke et al \cite{Vincke}\\
  $ 10^{12}$ G
               & 408.3894  & 17.1425    & 0.283  & Present\\
               & ---       & 17.0588    & 0.28   & Lai et al \cite{Salpeter:1992}\\
               & 408.566   & 16.966     & 0.278  & Wille \cite{Wille:1988}\\
  $ 1000$ a.u  & 977.2219  & 22.7781    & 0.220  & Present\\
               & ---       & 21.6688    & 0.219  & Wille \cite{Wille:1988}\\
               & ---       & 22.7069    & 0.221  & Wunner et al \cite{Wunner:82}\\
               & ---       & 22.67      & 0.222  & Larsen \cite{Larsen}\\
               & ---       & 22.7694    & 0.219  & Vincke et al \cite{Vincke}\\
  $ 10^{13}$ G
               & 4219.565  & 35.7539    & 0.147  & Present\\
               & 4231.82   & 23.52      & 0.125  & Wille \cite{Wille:1988}\\
               & ---       & 35.74      & 0.15   & Lai et al \cite{Salpeter:1992}\\
   $ 4.414\times{10}^{13} $ G
               & 18728.48  & 54.4992    & 0.101  & Present\\
\end{tabular}
\end{ruledtabular}
\end{table*}
\endgroup

\begin{table}
\caption{\label{Table:2} Total $E_T$, binding $E_b$ energies  and
    equilibrium distance  $R_{eq}$ for the $1_g$ state  at $\tha=45^{\circ}$.
    Optimal value of the gauge parameter $\xi$ is given (see text).}
\begin{ruledtabular}
\begin{tabular}{lcccc}
\( B \)& \( E_T \) (Ry) &\(    E_{b} \) (Ry) &\( R_{eq} \) (a.u.)& \(
\xi  \)\\[5pt] 
\hline
 \( 10^{9} \) G
              & -1.14248  & 1.56801  & 1.891 & 0.5806  \\[5pt]
 \( 1 \) a.u.
              & -0.918334 & 1.91833  & 1.667 & 0.5846  \\[5pt]
\( 10^{10}\) G
              & 1.26195  & 2.99337 & 1.103 & 0.5958   \\[5pt]
 \( 10 \) a.u.
              & 6.02330  & 3.97670 & 0.812 & 0.6044   \\[5pt]
 \( 10^{11}\) G
              & 36.15638  & 6.39681 & 0.466 & 0.6249   \\[5pt]
 \( 100 \) a.u.
              & 91.70480  & 8.29520 & 0.337 & 0.6424   \\[5pt]
 \( 10^{12}\) G
              & 413.2988  & 12.2331 & 0.198 & 0.6894   \\[5pt]
 \( 1000\) a.u.
              & 985.1956  & 14.8044 & 0.147 & 0.7151   \\[5pt]
 \( 10^{13}\) G
              & 4236.342  & 18.9769 & 0.097 & 0.8277   \\[5pt]
 \( 4.414\times{10}^{13} \) G
              & 18760.77  & 22.2047 & 0.073 & 0.9276   \\[5pt]
\end{tabular}
\end{ruledtabular}
\end{table}

\begin{table*}
\caption{\label{Table:3} Total $E_T$, binding $E_b$ energies and 
  equilibrium distance  $R_{eq}$ for the  $1_g$ state  in  
  perpendicular configuration, $\tha=90^{\circ}$. 
  Optimal value of the gauge parameter $\xi$ is given (see text).}
\begin{ruledtabular}
 \begin{tabular}{lccccl}
\(B\)&  \( E_T \) (Ry)& \(E_{b}\) (Ry)& \(R_{eq}\) (a.u.)& \(\xi\)&
\\ 
\hline 
\( 10^{9}\) G
              & -1.137342 & 1.56287  & 1.875  & 0.6380 & Present\\
              &           & 1.56384  & 1.879  &        & Wille
             \cite{Wille:1988}\\
  \( 1 \) a.u.
              & -0.89911  & 1.89911  & 1.636  & 0.6448 & Present\\
              & ---       & 1.8988   & 1.634  &        & Larsen \cite{Larsen}\\
              & -0.89774  & 1.8977   & 1.65   &        & Kappes et al
             \cite{Schmelcher}\\
  \( 10^{10}\) G
              &  1.36207  & 2.89324  & 1.059  & 0.6621 & Present\\
              &  ---      & 2.8992   & 1.067  &        & Wille
             \cite{Wille:1988} \\
  \( 10\) a.u.
              &  6.23170  & 3.76830  & 0.772  & 0.6752 & Present\\
              & \emph{---}& 3.7620   & 0.772  &        & Larsen \cite{Larsen}\\
  \(10^{11} \) G
              & 36.7687   & 5.78445  & 0.442  & 0.7064 & Present\\
              & ---       & 5.6818   & 0.428  &        & Wille
             \cite{Wille:1988}\\
  \( 100\) a.u.
              & 92.7346   & 7.26543  & 0.320  & 0.7329 & Present\\
              & ---       & 7.229    & 0.320  &        & Larsen
             \cite{Larsen}\\
  \( 10^{12}\) G
              & 415.5621  & 9.96986  & 0.196  & 0.8034 & Present\\
              & ---       & 4.558    & 0.148  &        & Wille
             \cite{Wille:1988}\\
  \( 1000\) a.u.
              & 988.3082  & 11.6918 & 0.151  & 0.8520  & Present\\
              & ---       & 11.58    & 0.1578 &        & Larsen
             \cite{Larsen}\\
  \(10^{13}\) G
              & 4241.470  & 13.8490 & 0.113  & 0.9359 & Present\\
  \( 4.414\times 10^{13} \) G
              & 18767.50  & 15.4700  & 0.0937 & 0.9795 & Present\\
\end{tabular}
\end{ruledtabular}
\end{table*}

A failure to adequately reproduce the asymptotic behavior of the wave
functions leads to a failure in the proper description of the
electronic cloud shrinking in transversal to magnetic line directions
which become more and more essential with magnetic field growth.  The
origin of this shrinking is the Lorentz force action.  The above
drawback can easily be fixed by modifying the Hylleraas function by
multiplication on the lowest Landau orbital.  Nevertheless, it is
worth emphasizing that it is quite surprising that a linear superposition
of  Hylleraas type functions, which form a natural basis for
the free-field case, still allows one to get high accuracy results for
magnetic fields $B \lesssim 10^{11}$ G. Implicitly, it indicates that
the ground state wave function for these magnetic fields does not deviate 
drastically from free-field behavior. It is obvious that taking
the above-mentioned modified Hylleraas functions in the analysis made by
Wille \cite{Wille:1988} would allow us to reach the same accuracies but
with a much smaller basis.  We can treat the results by Larsen
\cite{Larsen} as an explicit demonstration that an insertion of the
Landau orbitals (multiplied by a Gaussian in $z$ function) in the
trial function (hence not in a fully adequate way as  follows from our
recipe) together with an optimization of the gauge dependence (in a
different manner than what we propose) leads to rather accurate
results.  Finally, it is worth mentioning that in the domain of very
strong magnetic fields, $B \gtrsim 10^{12}$ G, our results are more
accurate than those obtained by the Thomas-Fermi method
\cite{Salpeter:1992}.

The results for $\tha=45^{\circ}$ are shown in Table II, where a
gradual shortening of the equilibrium distance is accompanied by an
increase of total and binding energies with magnetic field growth. It
is worth noting that the parameter $\xi$ evolves from about $0.5$ to
$0.93$ with magnetic field growth, thus changing from symmetric gauge
for weak fields to an almost asymmetric one for strong ones. This
phenomenon  takes place for all orientations $\tha \neq 0$,
becoming more and more pronounced with the inclination angle growth
(see below). We are unaware of any other calculations for
$\tha=45^{\circ}$ to compare with.

For the perpendicular configuration ($\tha = 90^{\circ}$) the results
are presented in Table III. Similar to what appeared for the parallel
configuration (see above) our results are again slightly less accurate
than those of Wille for $B \lesssim 10^{10}$ G becoming the most  
accurate for stronger fields. In particular, it indicates that a
domain of applicability of the trial function, taken in the form of
a superposition of Hylleraas type functions, reduces when the inclination
grows. The results reported by Larsen \cite{Larsen} and by
Kappes-Schmelcher \cite{Schmelcher} are slightly worse than ours
although the difference is very small.  The evolution of the gauge
parameter follows  a similar trend as was observed at $\tha =
45^{\circ}$, varying from $\xi=0.64$ to $\xi=0.98$ with magnetic field
growth from $B=10^9$ G to $B=4.414 \times 10^{13}$ G 
\footnote{$\xi=0.5$ at $B=0$}.

The total energy dependence of H$_2^+$ as a function of the
inclination angle $\tha$ for different magnetic fields is shown in
Fig.2.  The dotted line corresponds to the $H$-atom total energy in
the magnetic field.  For weak magnetic fields the hydrogen atom total
energy is higher than that of the $H_2^+$-ion one. However, for $B
\gtrsim 1.8 \times 10^{11}$ G the total energy of the $H$-atom becomes
lower than the total energy of the H$_2^+$-ion for some angles
$\tha_{cr} < \tha < 90^{\circ}$. It implies that one proton in the
$H_2^+$ system can go to infinity. Thus, it manifests the appearance
of a {\it dissociation} channel $H_2^+ \rar H + p$. At first, the
dissociation occurs at $\tha=90^{\circ}$. Afterward the domain of
inclinations with allowed dissociation widens with the magnetic field
growth reaching $25^{\circ} \lesssim \tha \leqslant 90^{\circ}$ for $B
= 4.414 \times 10^{13}$ G. The dependence of the critical angle
$\tha_{cr}$ on the magnetic field is shown in Fig.3.  Naively, it
looks like  the rate of dissociation is maximal at $\tha =
90^{\circ}$. However, a precise conclusion depends on the form of the barrier
or, in other words, on the form of the potential surface in $\tha, R$. The
rate of dissociation as a function of inclination angle and magnetic
field is not studied in detail in present work and will be published
elsewhere.

\begin{figure*}
  \begin{center}
    \[
    \begin{array}{cc}
    {\includegraphics[width=2.0in,angle=-90]{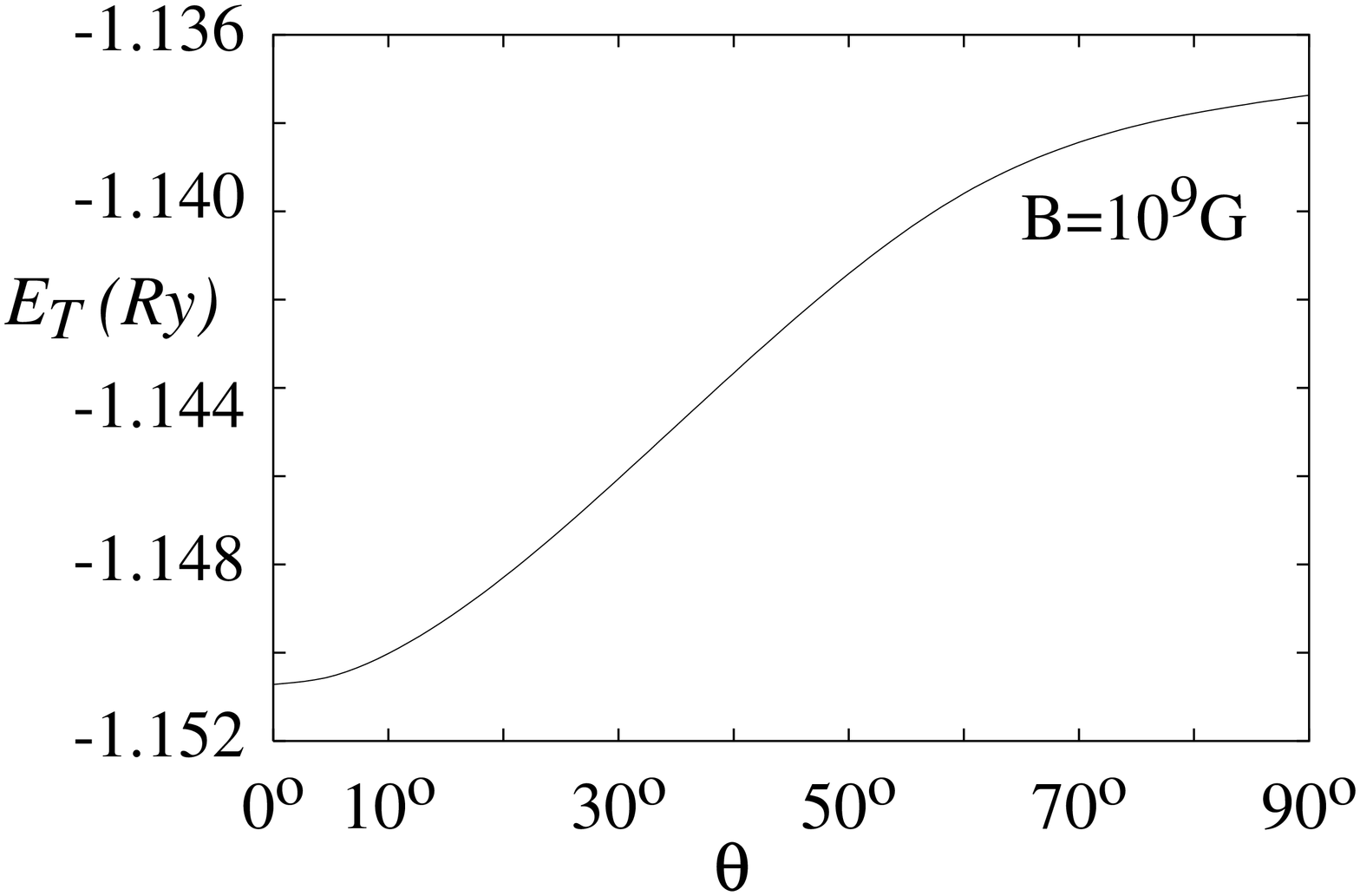}} &
    {\includegraphics[width=2.0in,angle=-90]{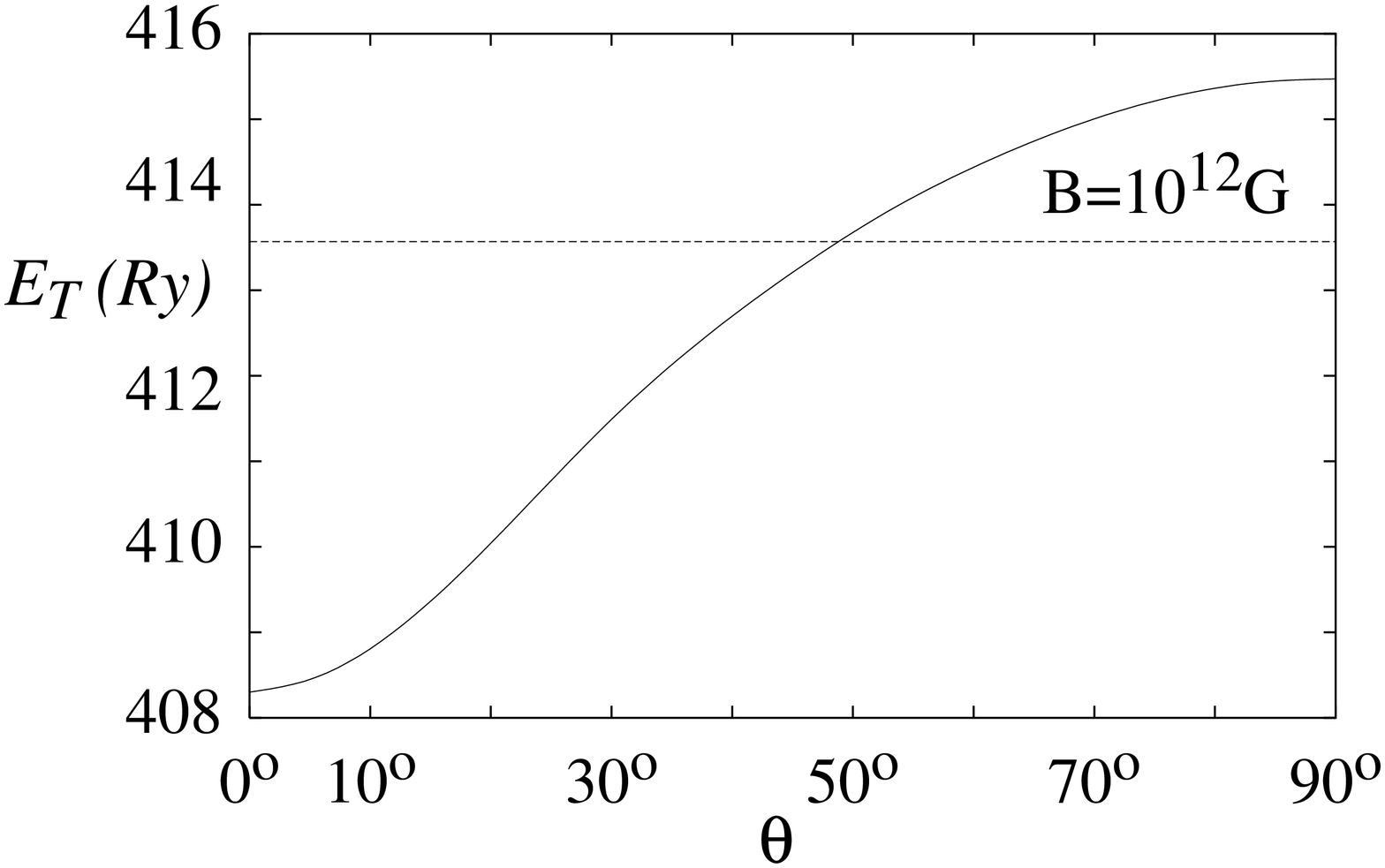}}     \\
    {\includegraphics[width=2.0in,angle=-90]{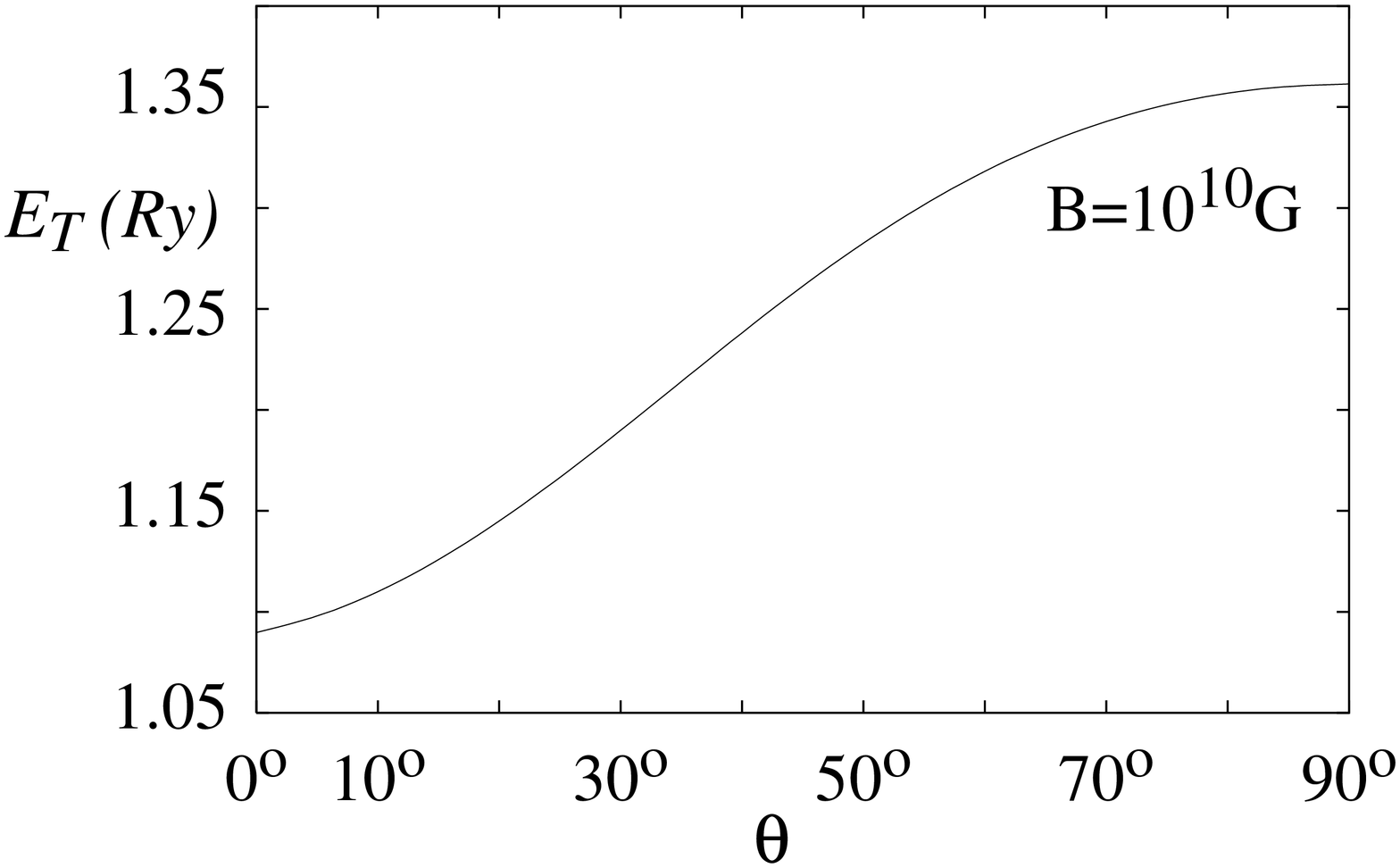}}&
    {\includegraphics[width=2.0in,angle=-90]{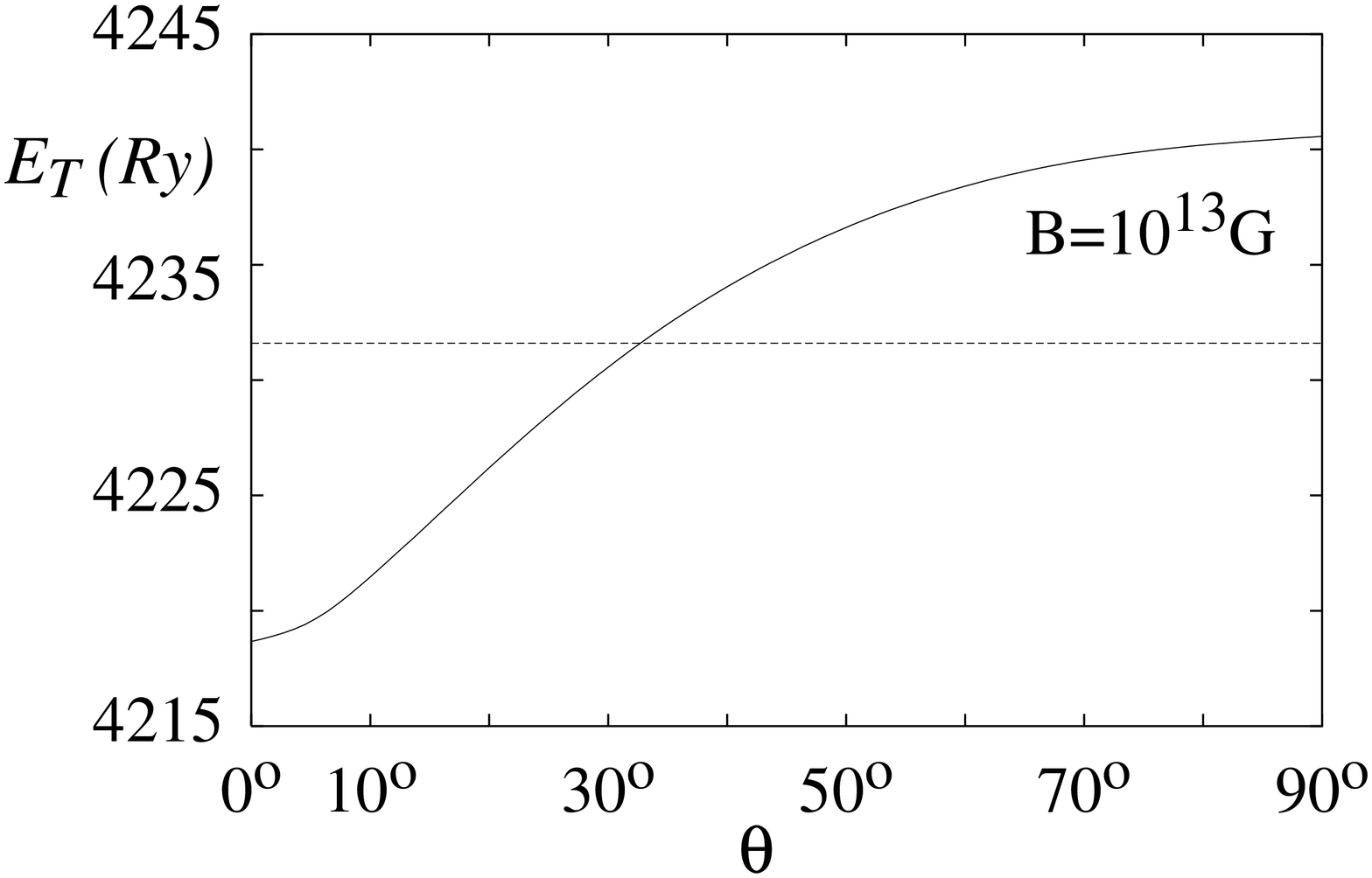}}     \\
    {\includegraphics[width=2.0in,angle=-90]{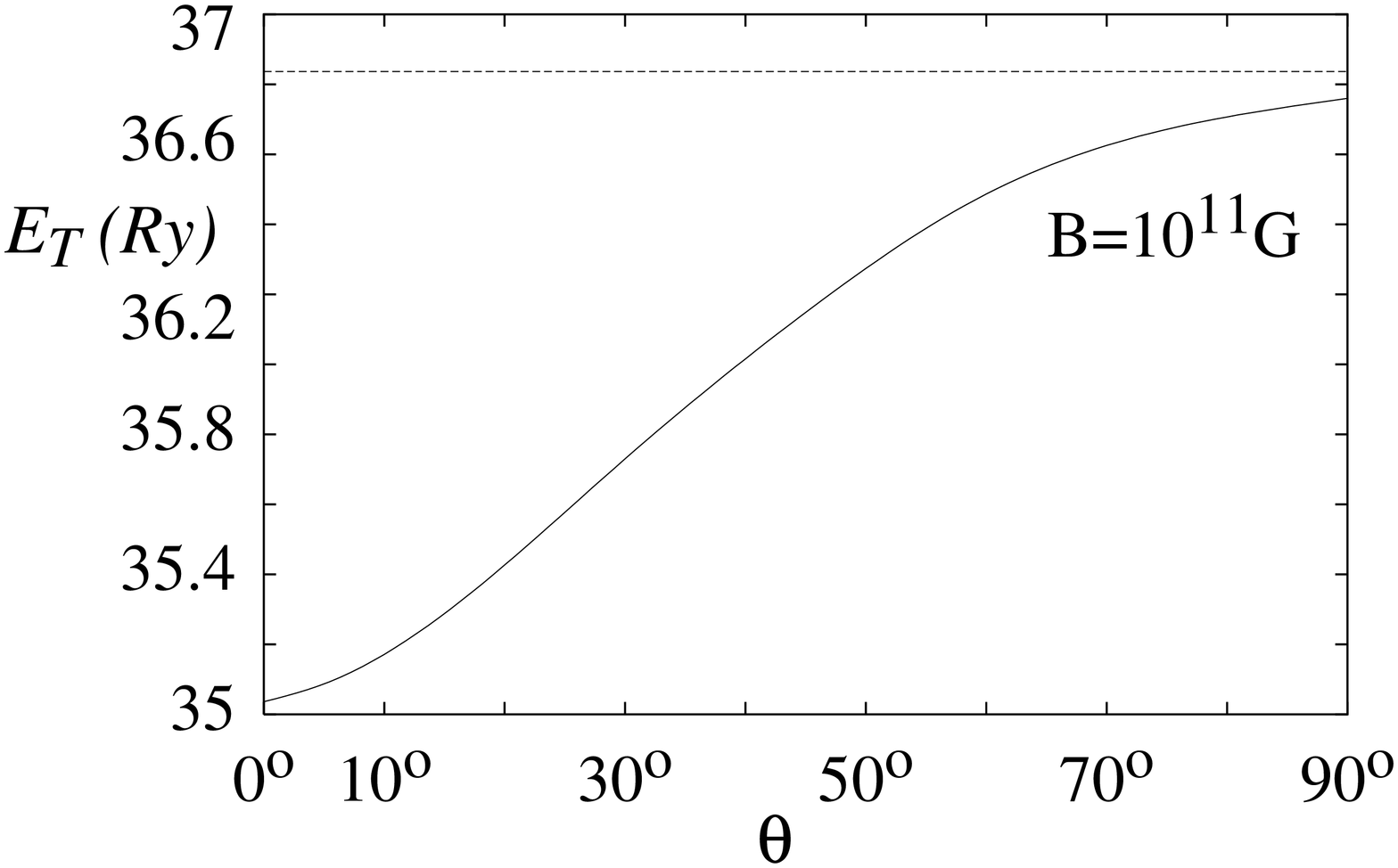}}&
    {\includegraphics[width=2.0in,angle=-90]{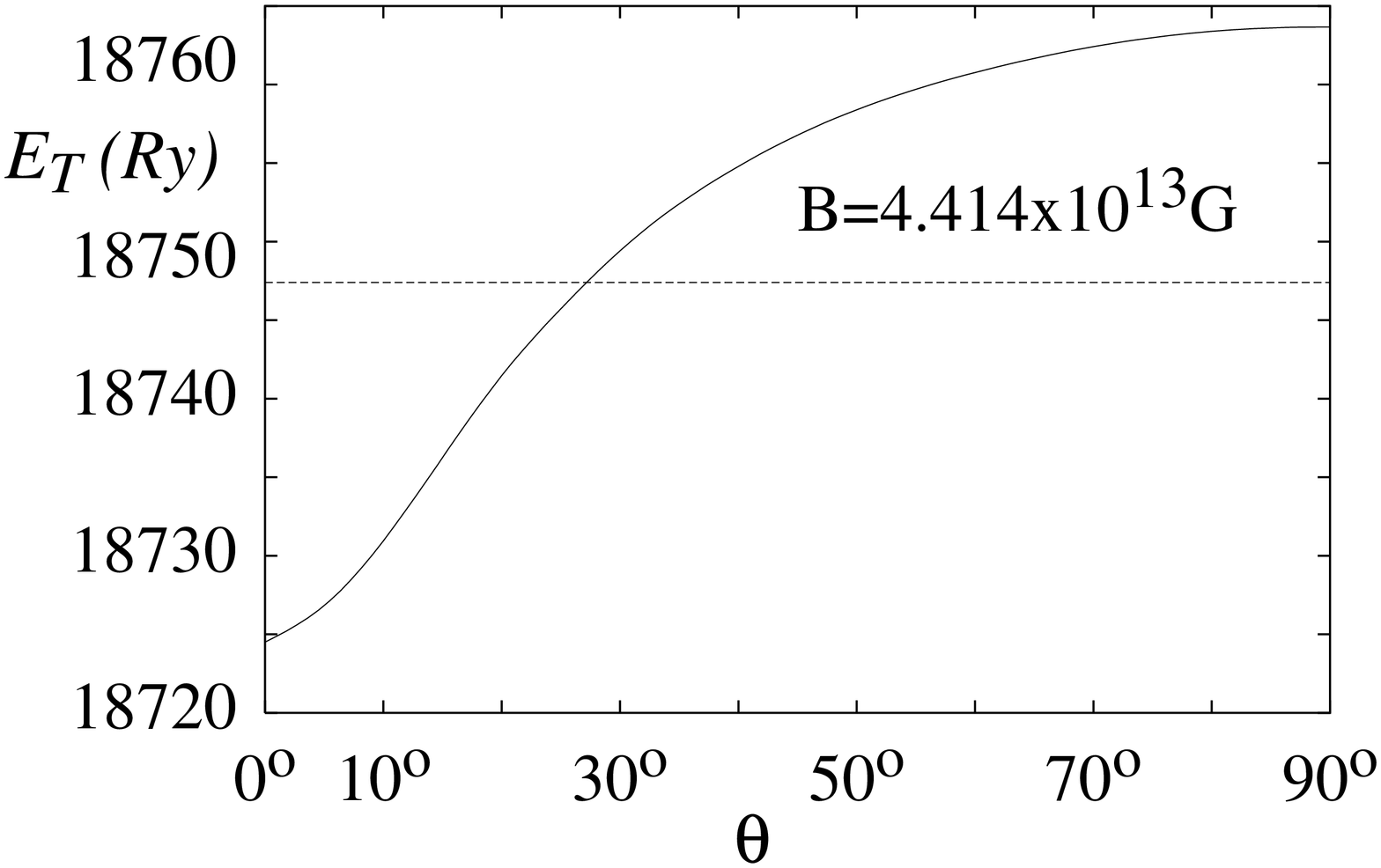}}
    \end{array}
     \]
    \caption{\label{fig:2} H$_2^+$ total energy ($E_T$) for the ground state
      $1_g$ as function of the inclination angle $\tha$ for different
      magnetic fields. The dotted lines correspond to the H-atom total
      energy taken from \cite{Salpeter:1992}.}
  \end{center}
\end{figure*}

\begin{figure}
\includegraphics*[width=2.7in,angle=-90]{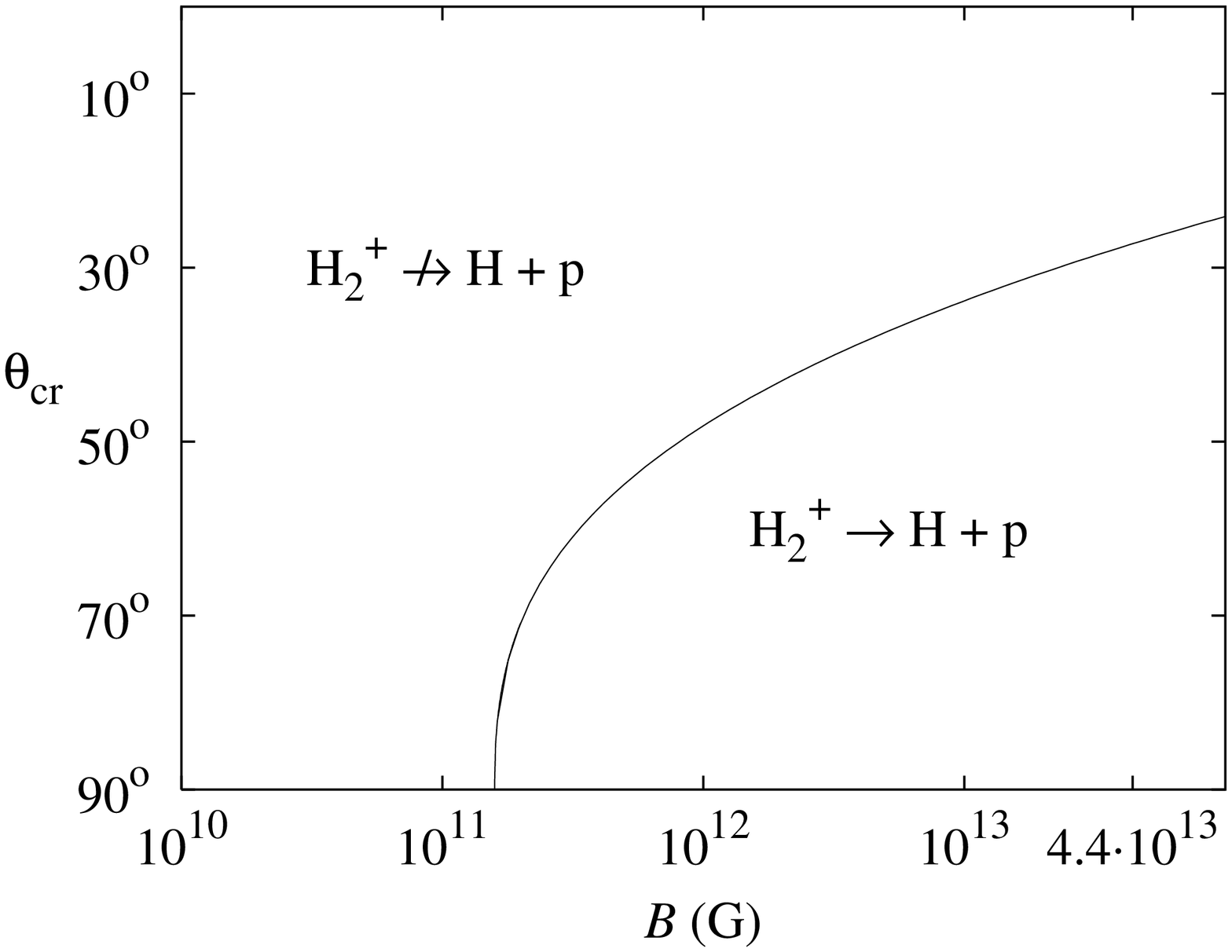}%
\caption{\label{fig:3} $H_2^+$: domains of dissociation $\leftrightarrow$
      non-dissociation for the $1_g$ state.}
\end{figure}

We observe that for fixed value of the magnetic field strength $B$ the
binding energy $E_b$ as a function of $\tha$ always decreases when
changing from the parallel to the perpendicular configuration (see
Fig.2). A similar picture holds for all studied values of the magnetic
field strength. Thus, we can draw the conclusion that the molecular
ion becomes less and less stable monotonically as a function of
inclination angle growth.  It confirms the statement \cite{Kher,
  Wille:1988, Larsen, Schmelcher}, that the {\it highest molecular
  stability of the $1_g$ state occurs for the parallel configuration}. We
extend its validity to magnetic field strengths $B \lesssim 4.414
\times 10^{13}$ G. It is worth  emphasizing that the rate of increase
of binding energy with magnetic field growth depends on the inclination --
it slows down with inclination increase. This effect means that
$H_2^+$ in the parallel configuration becomes more and more stable towards
rotations -- the energy of the lowest rotational state should 
increase rapidly with magnetic field (see Table V and discussion there).

Regarding the internuclear equilibrium distance $R_{eq}$, one would
straightforwardly expect that it will always decrease with inclination
growth. Indeed, for all studied magnetic fields we observe that
$R_{eq}$ at $\tha=0^{\circ}$ is larger than $R_{eq}$ at
$\tha=90^{\circ}$ (cf.  Tables I,III). This can be explained as a
consequence of the much more drastic shrinking of the electronic cloud
in the direction transverse to the magnetic field than in the
longitudinal one.  Actually, for magnetic fields $B \lesssim 10^{12}$
G the equilibrium distance $R_{eq}$ decreases monotonically with
inclination growth, as seen in Fig.4. However, this trend breaks down
for higher magnetic fields where the shortest equilibrium distances
occur for orientations $\tha_{min} < 90^{\circ}$ (!).  Furthermore, as
the magnetic field grows the molecular ion becomes the most compact
for smaller angles, being $\tha_{min} \sim 60^{\circ}$ for $B \lesssim
10^{12}$ G and then going down to $\tha_{min} \sim 30^{\circ}$ for $B
= 4.414\times 10^{13}$ G.  The minimal value of $R_{eq}$ also deepens
in comparison with $R_{eq}$ at $\tha=90^{\circ}$. For example, for $B
= 4.414\times 10^{13}$ G it becomes almost twice as small as $R_{eq}$
at $\tha=90^{\circ}$ . All these irregularities appear when at the
same time the binding energy decreases monotonically as inclination
grows (see Fig.2). We do {\it not} have a physical explanation of
this phenomenon yet.

\begin{figure*}
  \begin{center}
    \[
    \begin{array}{cc}
     \includegraphics[width=2.0in,angle=-90]{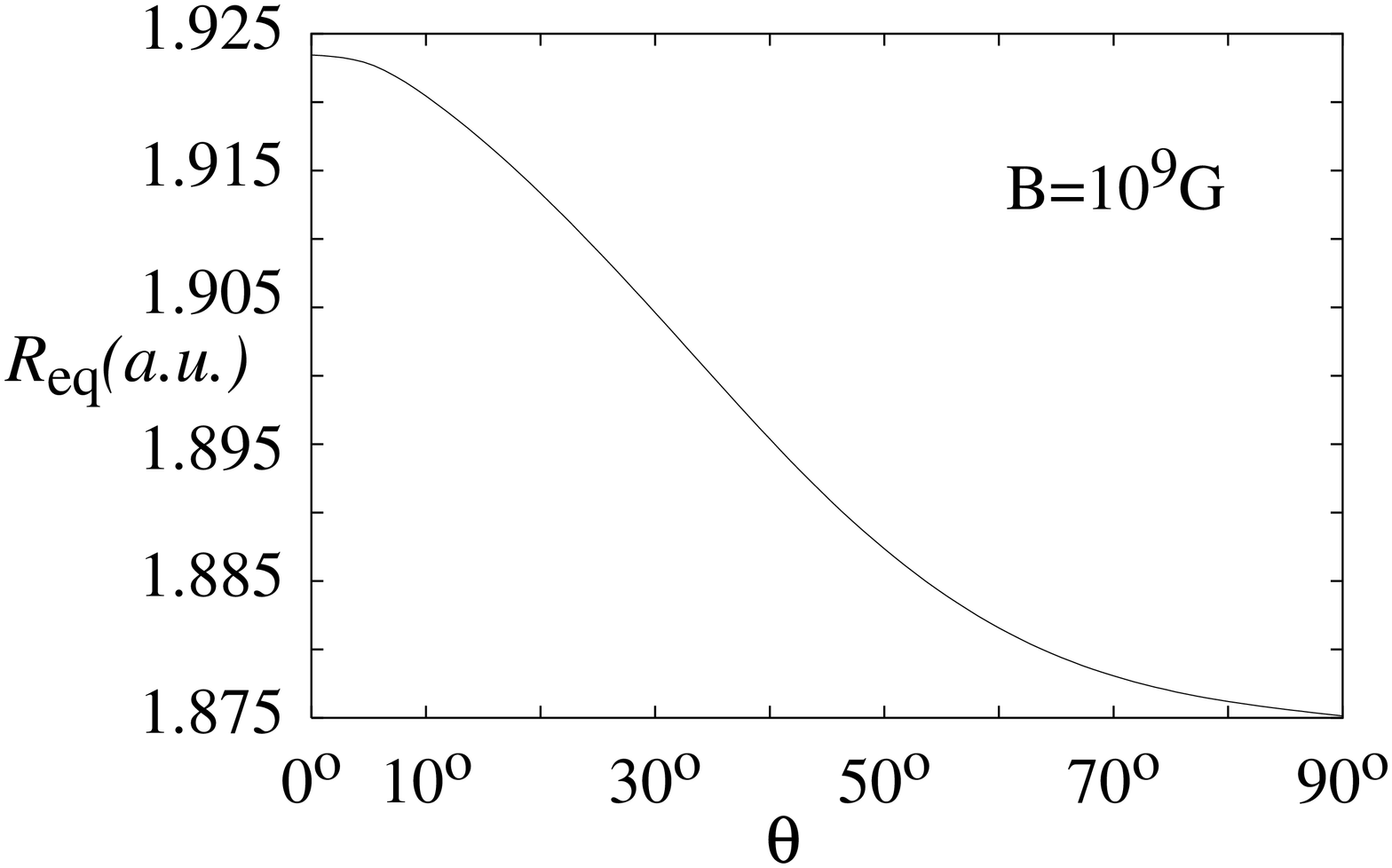} &
     \includegraphics[width=2.0in,angle=-90]{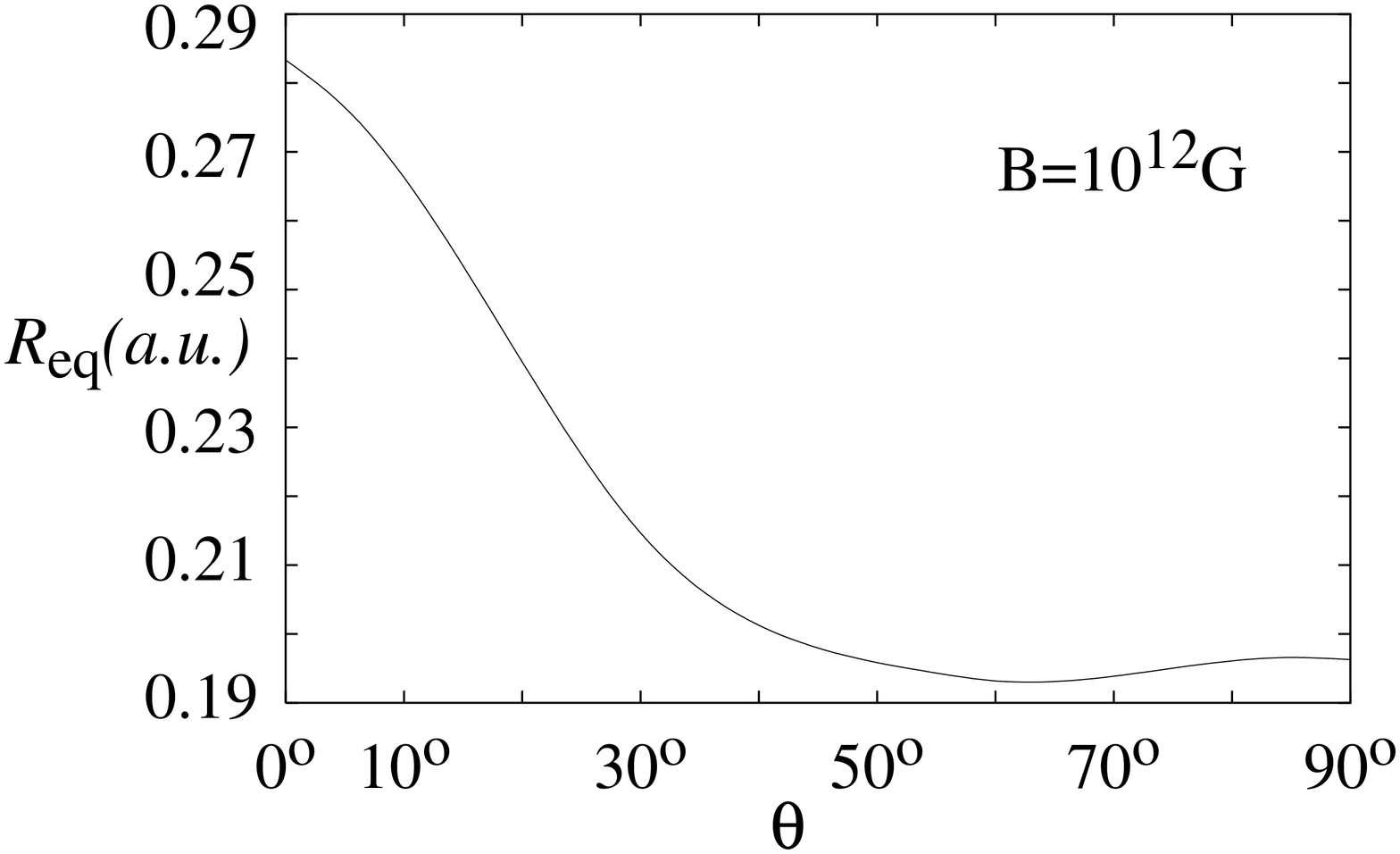} \\
     \includegraphics[width=2.0in,angle=-90]{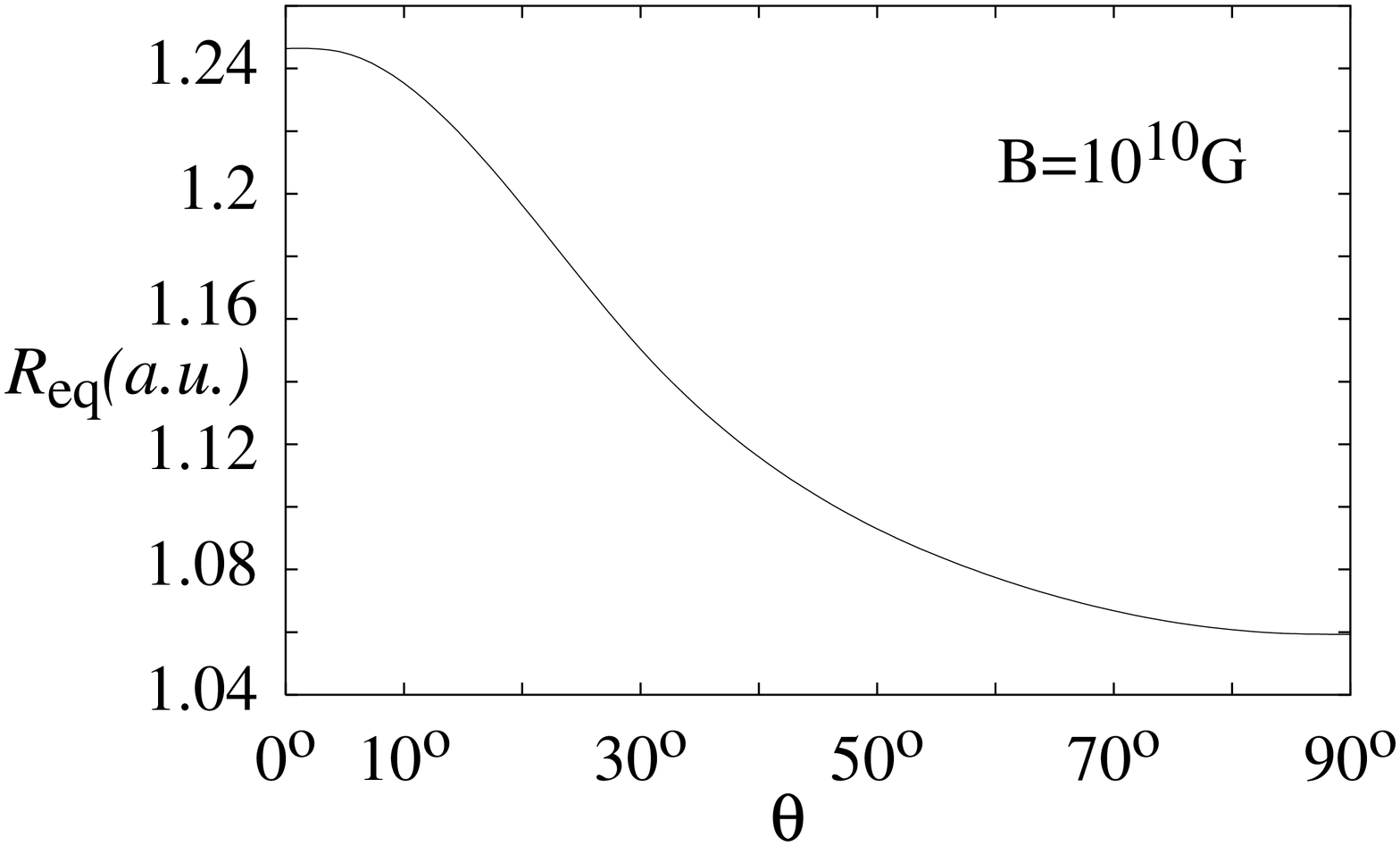}&
     \includegraphics[width=2.0in,angle=-90]{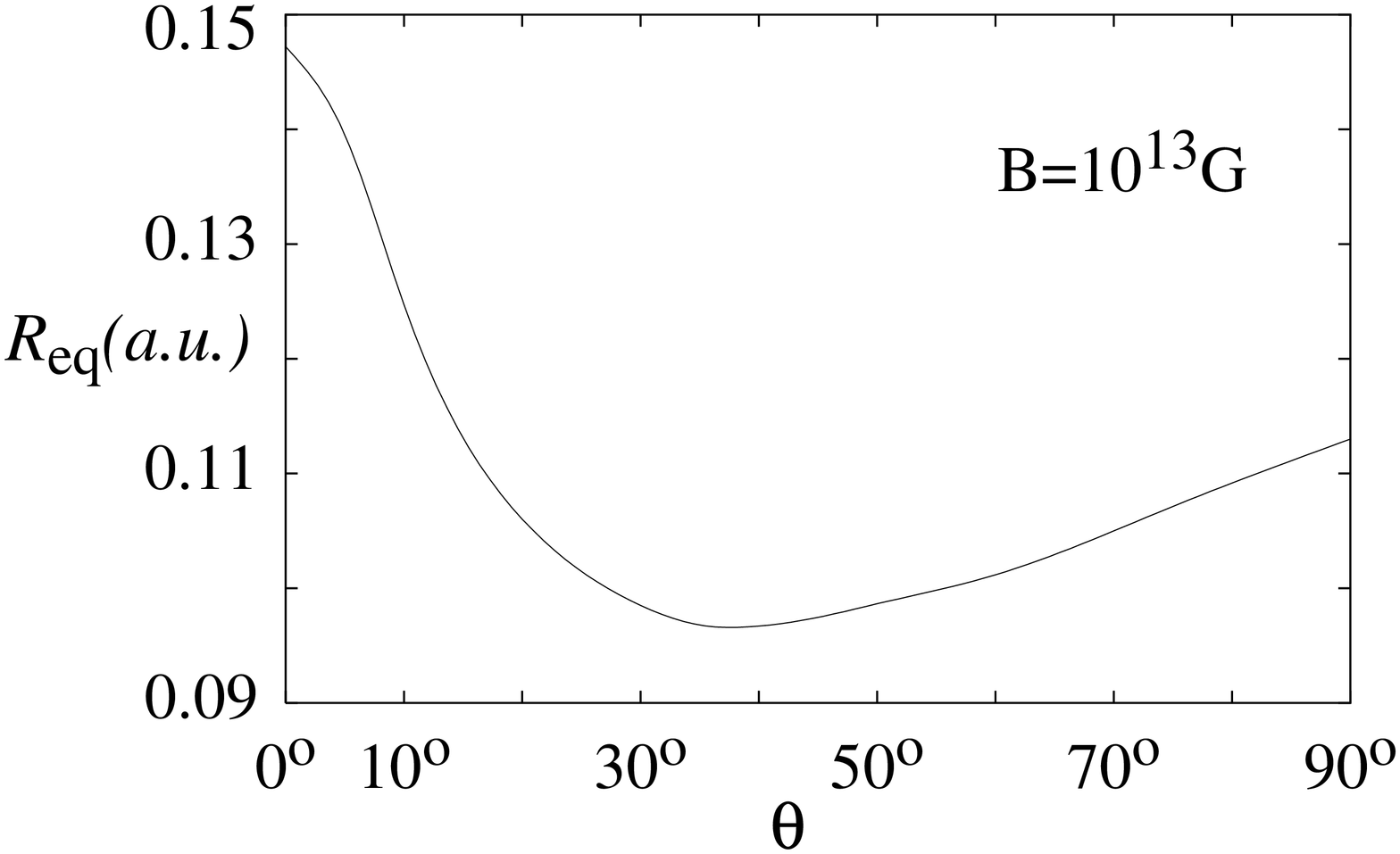}\\
     \includegraphics[width=2.0in,angle=-90]{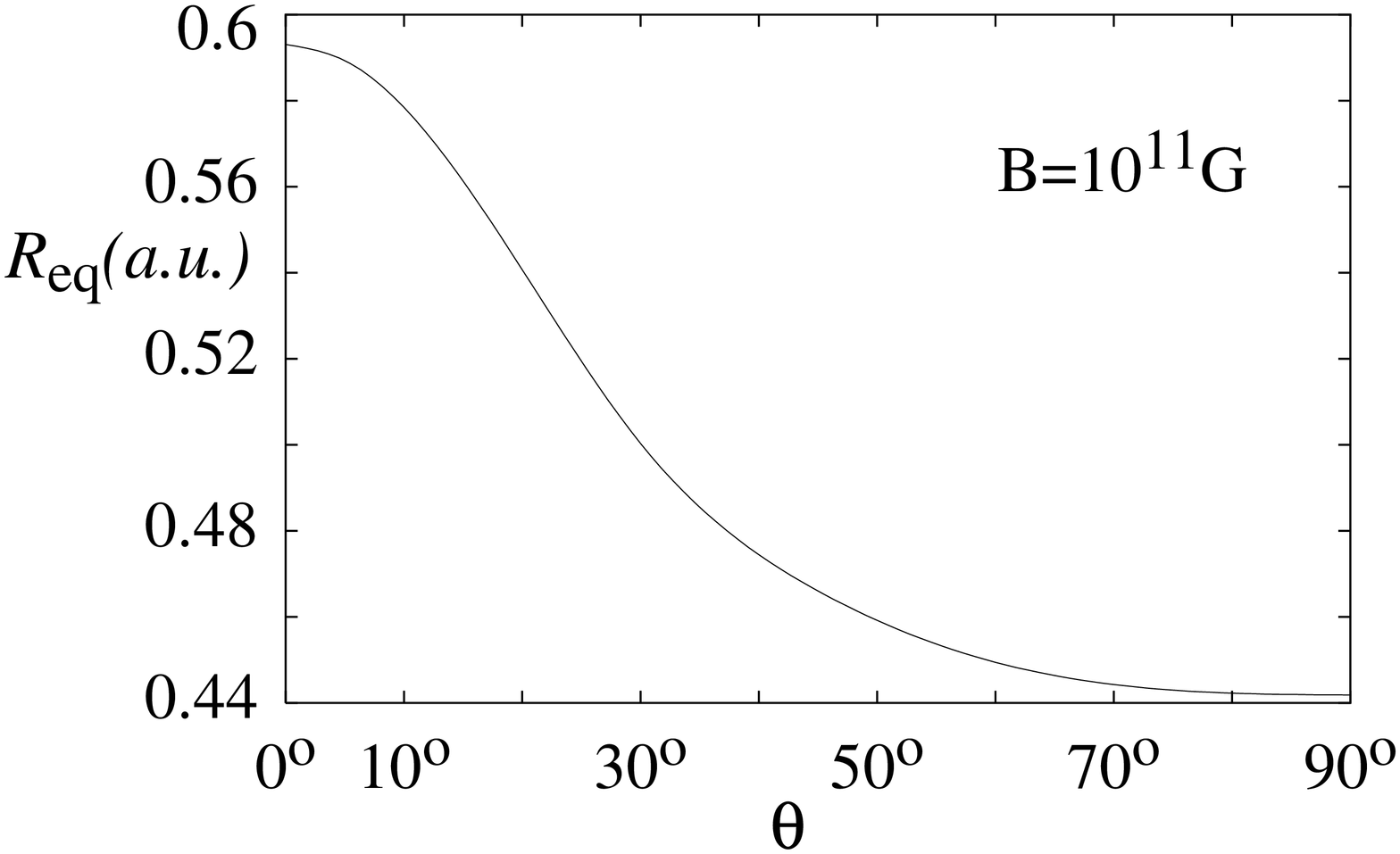}&
     \includegraphics[width=2.0in,angle=-90]{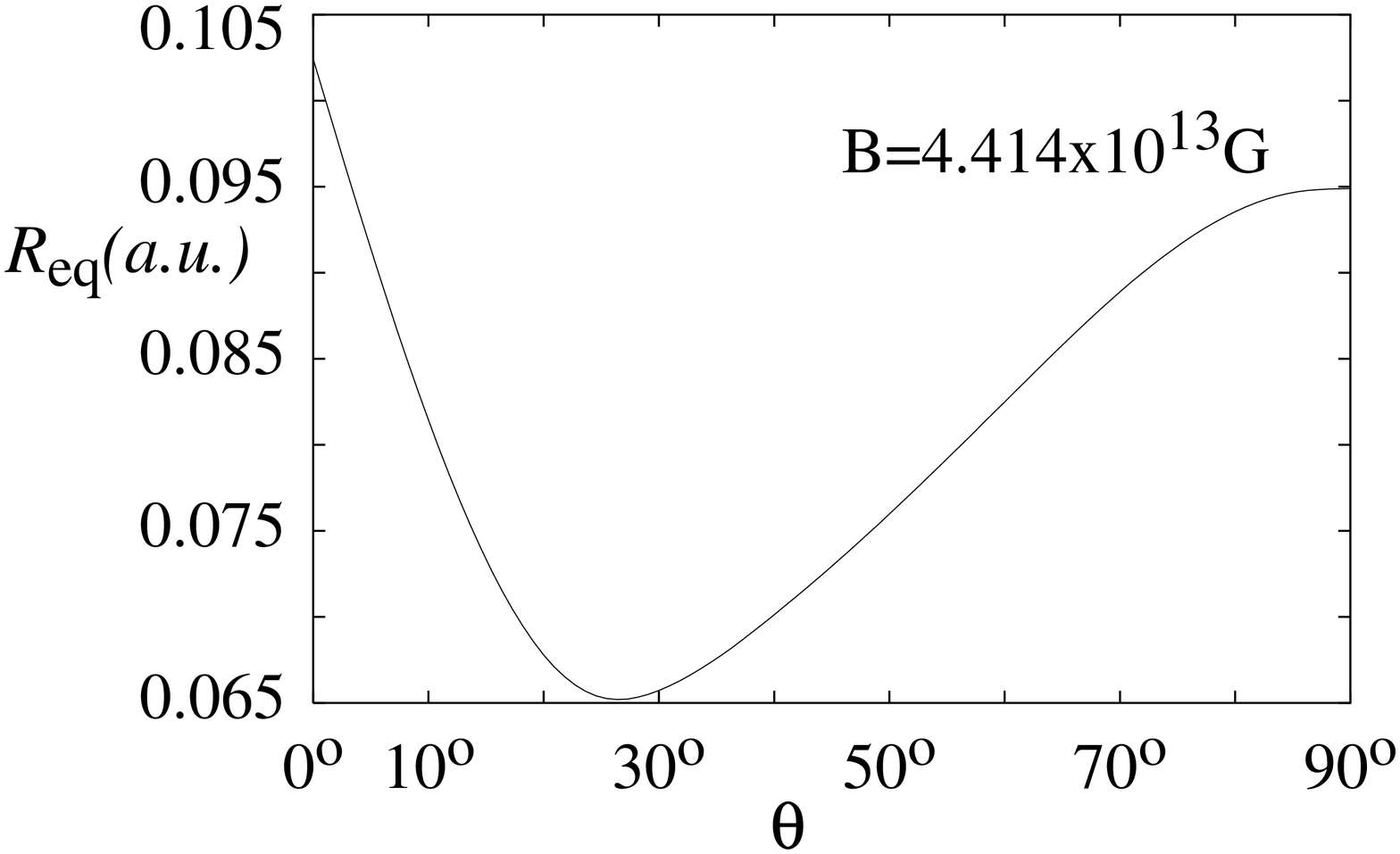}
    \end{array}
     \]
    \caption{\label{fig:4} $H_2^+$ equilibrium  distance as function of the
      inclination angle $\tha$ for the  $1_g$ state. }
  \end{center}
\end{figure*}

\begin{figure}
     {\includegraphics*[width=4.0in,angle=-90]{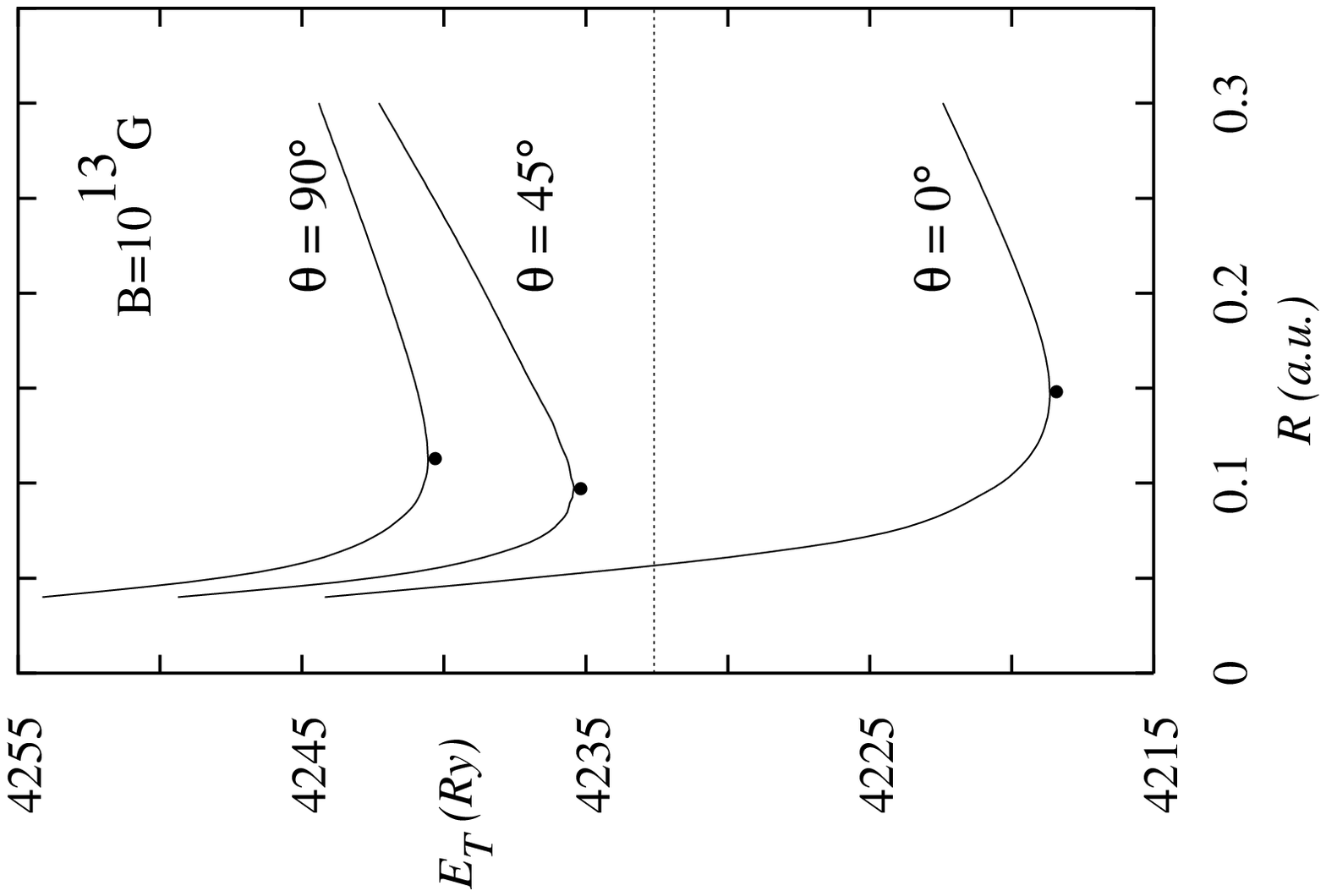}}
    \caption{\label{fig:5} $H_2^+$ potential curves (total energy 
      $E_T$ vs $R$) 
      for the $1_g$ state at $B=10^{13}$ G and different inclinations.
      The dotted line corresponds to the H-atom total energy taken
      from \cite{Salpeter:1992}. Positions of minimum are marked by bullets.
      }
\end{figure}

Fig.5 illustrates the above-mentioned non-monotonic behaviour of the
internuclear distances near the equilibrium position, $R \lesssim 0.3$
a.u., for different orientations at $B=10^{13}$ G. For all values of
$\tha$ a clear minimum in $R$ develops. For certain orientations
($\tha \gtrsim 45^{\circ}$) the potential energy curves are situated
above the $H$-atom total energy. The total energy curves lie at
increasingly higher energies as inclination grows. Hence, one can draw
the conclusion that the growth of the inclination angle leads to an
excitation of the $H_2^+$ system. It is quite interesting to make a
comparison of the present situation with what appears in chemistry.
The typical situation for the behaviour of molecular electronic
excitations of standard chemical systems (in the absence of a magnetic
field) is characterized by an increase of equilibrium distance with
energy growth. This is explained either by enhancement of the
antibonding character or suppression of the bonding character. Our
situation is the opposite -- in general the antibonding character is
suppressed or, equivalently, the bonding character is enhanced.
However, for strong magnetic fields $B > 10^{12}$ G, where an abnormal
behavior of the equilibrium distance is seen (see Fig.4), the
situation is reminiscent of the chemical one.  Starting from
$\tha_{cr}$ the growth of inclination leads to an increase in the
equilibrium distance.  We do not know if this analogy makes physical
sense or it is simply a coincidence.

\begin{table}
\caption{ \label{Table:4} $1_g$ state: Expectation values of the transversal
  $<\rho>$ and longitudinal $<|z|>$ sizes of the electron
  distribution in H$_2^+$ in a.u. at different orientations
  and magnetic field strengths.  At $\tha = 0^{\circ}$ the expectation
  value $<\rho>$ almost  coincides to the cyclotron radius of electron.}
\begin{ruledtabular}
\begin{tabular}{lcccccc}
\( B \) & \multicolumn{3}{c}{\( <\rho > \) }&
\multicolumn{3}{c}{ \( <|z|> \) }\\ & \( 0^{o} \)& \( 45^{o} \)&
\( 90^{o} \)& \( 0^{o} \)& \( 45^{o} \)& \( 90^{o} \)\\ 
\hline 
\( 10^{9} \) G& 0.909 & 1.002&
 1.084 &
1.666& 1.440 &
 1.180 \\
\( 1 \) a.u.& 0.801 & 0.866 & 0.929 &
 1.534 &
1.313 & 1.090\\ 
\( 10^{10} \) G& 0.511 & 0.538 & 0.569 &
1.144 &
 0.972 &
0.848\\ 
\( 10 \) a.u.& 0.359 & 0.375 & 0.396 & 0.918& 0.787
& 0.708\\ 
\( 10^{11} \) G& 0.185 & 0.193 & 0.205 & 0.624 &
0.542 & 0.514\\ 
\( 10 \)0 a.u.& 0.123& 0.129& 0.139 &
0.499& 0.443 &
 0.431 \\
\( 10^{12} \) G&
 0.060 &
0.065 &
 0.074 &
 0.351 &
0.324 & 0.340\\ 
\( 10 \)00 a.u.& 0.039 & 0.043 & 0.054 &
 0.289 &
0.275 & 0.290 \\ 
\( 10^{13} \) G& 0.019 & 0.025 & 0.037&
0.215 &
 0.221 &
0.256\\ 
\( 4.414\times 10^{13} \) G& 0.009&
 0.017 &
0.030 & 0.164 & 0.191 & 0.232 \\ 
\end{tabular}
\end{ruledtabular}
\end{table}

In order to characterize the electronic distribution for different
orientations we have calculated the expectation values of the
transversal $<\rho>$ and longitudinal $<|z|>$ sizes of the electronic
cloud (see Table IV). The ratio
\[
\frac{<\rho>}{<|z|>} \ < \ 1\ ,
\]
quickly decreases with magnetic field growth, especially for small
inclination angles. It reflects the fact that the electronic cloud has
a more and more pronounced needle-like form, oriented along the
magnetic line as was predicted in
[\onlinecite{Kadomtsev:1971}-\onlinecite{Ruderman:1971}]. The
behaviour of $<\rho>$ itself does not display any unusual properties,
smoothly decreasing with magnetic field, quickly approaching the
cyclotron radius for small inclinations at large magnetic fields. On
the contrary, the $<|z|>$ behaviour reveals some surprising features. At
the beginning, at small magnetic fields, the $<|z|>$ expectation value
monotonically  decreases with inclination, but then after some irregular
behavior at $10^{12} \lesssim B \lesssim 10^{13}$, it begins a
monotonic increase with inclination. It is quite striking that there
is a domain of magnetic fields where $<|z|>$ has almost no dependence
on inclination.

As a result of our analysis the parallel configuration turned out to
be optimal for all studied magnetic fields. Therefore, it makes sense
to perform a study of the lowest vibrational state and also the lowest
rotational state (see Table V). In order to do this we separate the
nuclear motion along the molecular axis near equilibrium in the
parallel configuration (vibrational motion) and deviation in $\tha$ of
the molecular axis from $\tha=0^{\circ}$ (rotational motion).  The
vicinity of the minimum of the potential surface $E(\tha, R)$ at $\tha
= 0^{\circ}, R=R_{eq}$ is approximated by a quadratic potential and
hence we arrive at a two-dimensional harmonic oscillator problem in
$(R, \tha)$.  Corresponding curvatures near the minimum define the
vibrational and rotational energies (for precise definitions and
discussion see, for example, \cite{Larsen}). We did not carry out a
detailed numerical analysis, making rough estimates of the order of
$20 \%$.  For example, at B$=10^{12}$ G we obtain $E_{vib}= 0.276$ Ry
in comparison with $E_{vib}=0.259$ Ry given in \cite{Lopez-Tur:2000},
where a detailed variational analysis of the potential electronic
curves was performed.  Our estimates for the energy, $E_{vib}$, of the
lowest vibrational state are in reasonable agreement with previous
studies.  In particular, we confirm a general trend of the
considerable increase of vibrational frequency {\it viz.} growth of
$B$ indicated for the first time by Larsen \cite{Larsen}. The energy
dependence on the magnetic field is much more pronounced for the
lowest rotational state -- it grows much faster than the vibrational
one with magnetic field increase. It implies that the $H_2^+$ in
parallel configuration becomes more stable for larger magnetic fields.
From a quantitative point of view the results obtained by different
authors are not in good agreement. It is worth mentioning that our
results agree for large magnetic fields $\gtrsim 10 a.u.$ with results
by Le Guillou et al. \cite{Legui}, obtained in the framework of the so
called `improved static approximation', but deviate drastically at $B=1
a.u.$, being quite close to the results of Larsen \cite{Larsen} and
Wille \cite{Wille:1987}. As for the energy of the lowest rotational
state our results are in good agreement with those obtained by other
authors (see Table V).

\begingroup
\squeezetable
\begin{table*}
  \caption{\label{Table:5}
    Energies of the lowest vibrational $(E_{vib})$ and rotational
    $(E_{rot})$ electronic states  associated with $1_g$ state
    at $\tha=0^{\circ}$.  The indexes in Le Guillou et al 
    \cite{Legui} correspond to the `improved adiabatic approximation'
    (a), and to the `improved static approximation' (b). }
\begin{ruledtabular}
\begin{tabular}{lcccl}
\( B \)& \( E_T \) (Ry)& \( E_{vib} \) (Ry)& \( E_{rot} \) (Ry)& \\ 
\hline
 \( 10^{9} \) G   & -1.15070 & 0.013 & 0.0053 & Present \\
                  &  ---     & 0.011 & 0.0038 & Wille
                  \cite{Wille:1987} \\
\( 1 \)a.u.       & -0.94991 & 0.015 & 0.0110 & Present \\
                  &    ---   & ---   & 0.0086 & Wille
                  \cite{Wille:1987} \\
                  &    ---   & 0.014 & 0.0091 & Larsen
                  \cite{Larsen} \\
                  &    ---   & 0.013 & ---    & Le Guillou et al (a)
                  \cite{Legui}\\
                  &    ---   & 0.014 & 0.0238 & Le Guillou et al (b)
                  \cite{Legui}\\
\( 10^{10} \) G      & 1.09044  & 0.028 & 0.0408 & Present \\
                     &   ---    & 0.026 & 0.0308 & Wille \cite{Wille:1987} \\
\( 10 \) a.u  & 5.65024  & 0.045 & 0.0790 & Present \\
              &    ---   & 0.040 & 0.133  & Larsen\cite{Larsen} \\
              &    ---   & 0.039 &  ---   & Le Guillou et al (a)
              \cite{Legui}\\
              &    ---   & 0.040 & 0.0844 & Le Guillou et al (b)
              \cite{Legui}\\
\( 10^{11} \) G      & 35.0434  & 0.087 & 0.2151 & Present \\
\( 100 \) a.u & 89.7096  & 0.133 & 0.4128 & Present \\
              &  ---     & 0.141 & 0.365  & Larsen\cite{Larsen}\\
              &  ---     & 0.13  &  ---   & Wunner et al \cite{Wunner:82}\\
              &    ---   & 0.128 &  ---   & Le Guillou et al (a)
              \cite{Legui}\\
              &    ---   & 0.132 & 0.410  & Le Guillou et al (b)
              \cite{Legui}\\
\( 10^{12} \) G      & 408.389  & 0.276 & 1.0926 & Present \\
                     &   ---    & 0.198 & 1.0375 & Khersonskij
                     \cite{Kher3} \\
\( 1000 \) a.u & 977.222  & 0.402 & 1.9273 & Present \\
               &   ---    & 0.38  & 1.77   &  Larsen\cite{Larsen} \\
               &   ---    & 0.39  &  ---   & Wunner et al \cite{Wunner:82}\\
               &    ---   & 0.366 &  ---   & Le Guillou et al (a)
               \cite{Legui}\\
               &    ---   & 0.388 & 1.916  & Le Guillou et al (b)
               \cite{Legui}\\
\( 10^{13} \) G      & 4219.565 & 0.717 & 4.875  & Present \\
                     &   ---    & 0.592 & 6.890  & Khersonskij
                     \cite{Kher3}  \\
\( 4.414\times 10^{13} \)G & 18728.48 & 1.249 & 12.065 & Present\\
\end{tabular}
\end{ruledtabular}
\end{table*}
\endgroup

We show the electronic distributions $|\psi(x,z)|^2$, integrated over
$y$ and normalized to one for magnetic fields $10^{9}, 10^{10},
10^{11}, 10^{12}$ G and different orientations in Fig.6. It was
already found \cite{Lopez:1997} that there is a change from ionic
(two-peak electronic distribution) to covalent coupling 
(single-peak distribution) at $\tha=0$. If for $B=10^9$ G, all
electronic distributions are characterized by two peaks, then for
$10^{12}$ G all distributions have a sharp single peak.  Fig.6
demonstrates also how the change of the type of coupling appears for
different inclinations. It is quite natural that for the perpendicular
configuration $\tha=90^{\circ}$, where the equilibrium distance is the
smallest, this change appears for smaller magnetic field.

\begin{figure*}
  \begin{center}
    \[
    \begin{array}{ccc}
    &   B=10^{9} \textrm{G} & \\
    \includegraphics[width=1.2in,angle=-90]{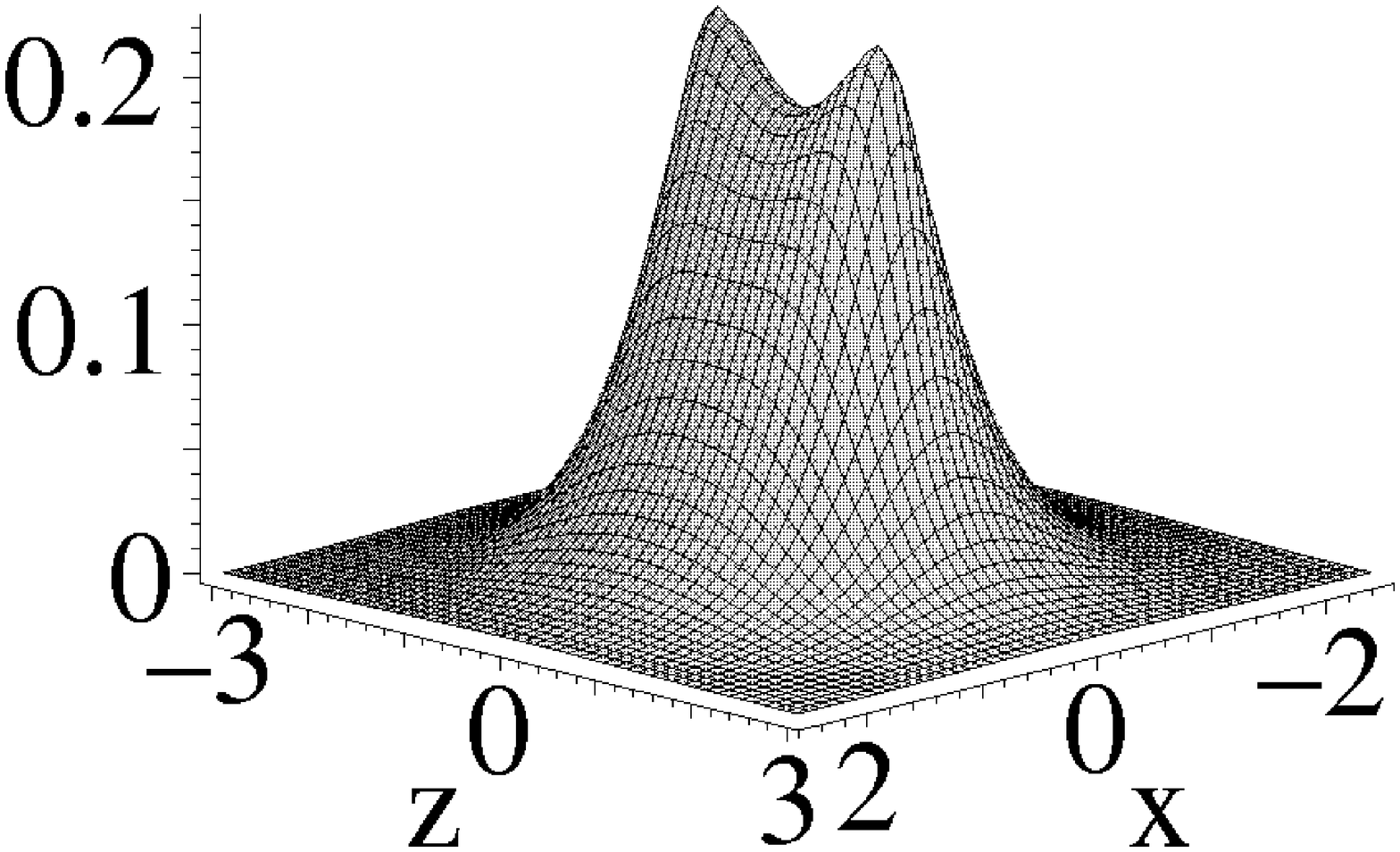}&
    \includegraphics[width=1.2in,angle=-90]{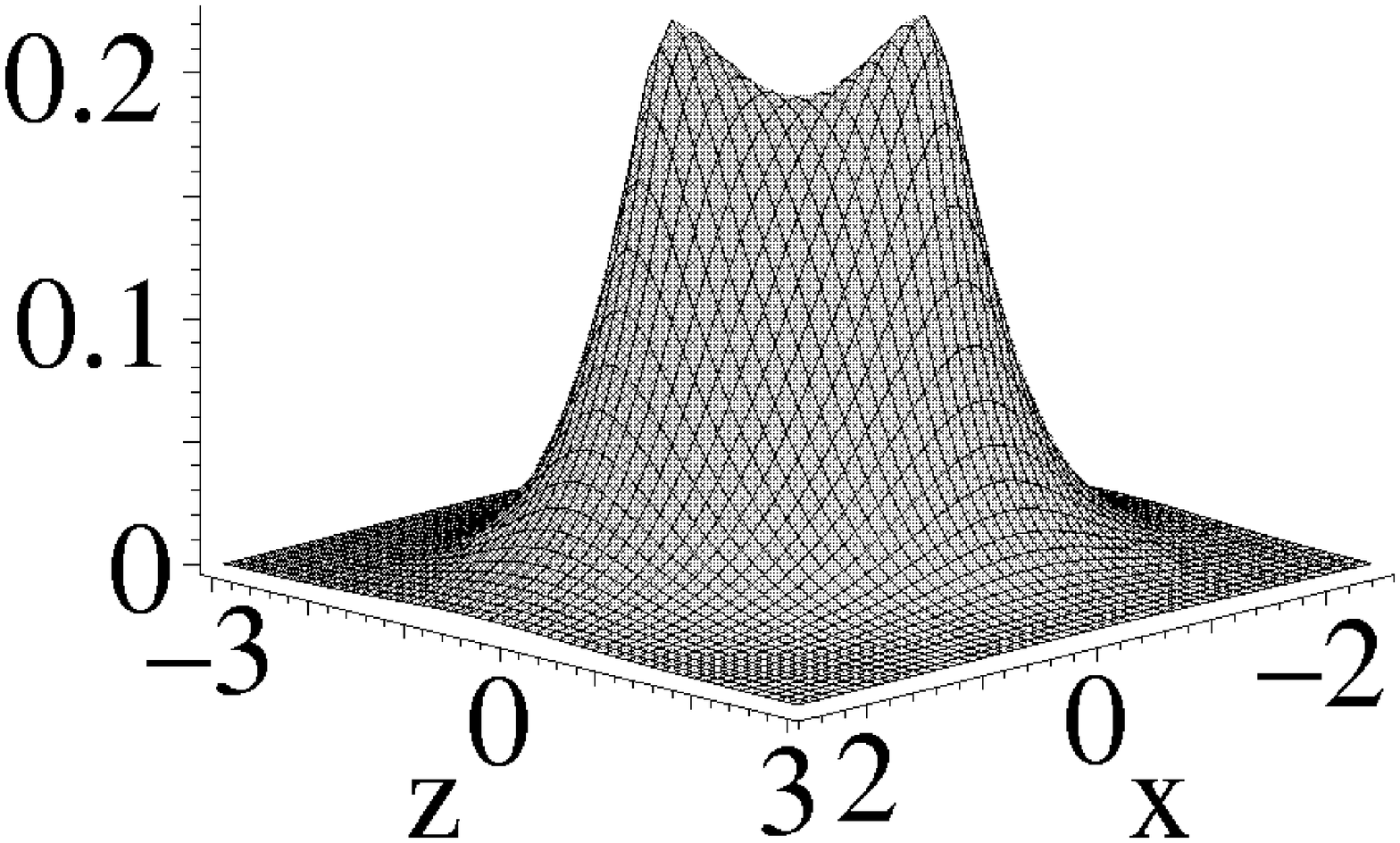}&
    \includegraphics[width=1.2in,angle=-90]{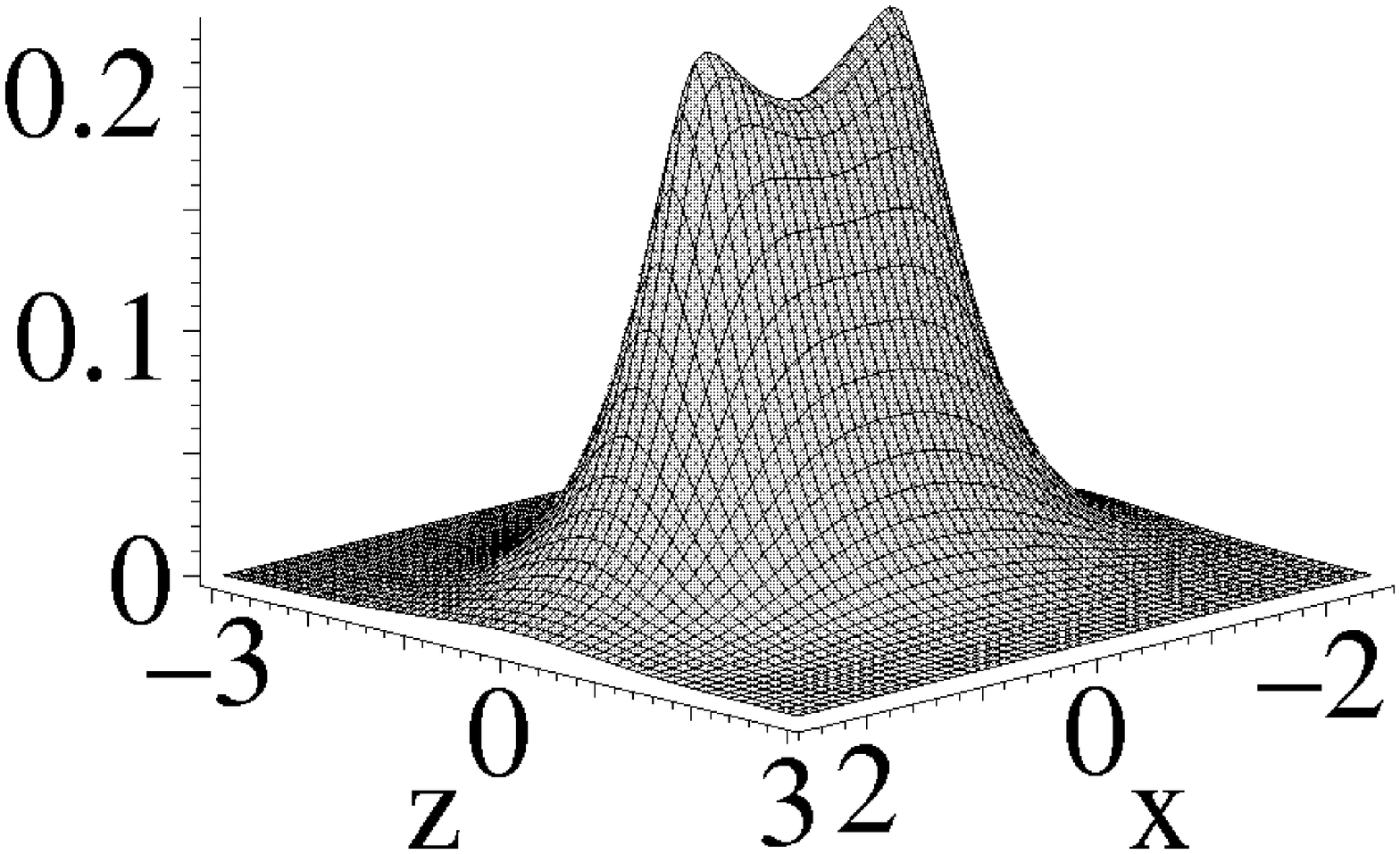} \\[5pt]
     \theta=0^{\circ} & \theta=45^{\circ} & \theta=90^{\circ}\\[15pt]
    & B=10^{10} \textrm{G} & \\
    \includegraphics[width=1.2in,angle=-90]{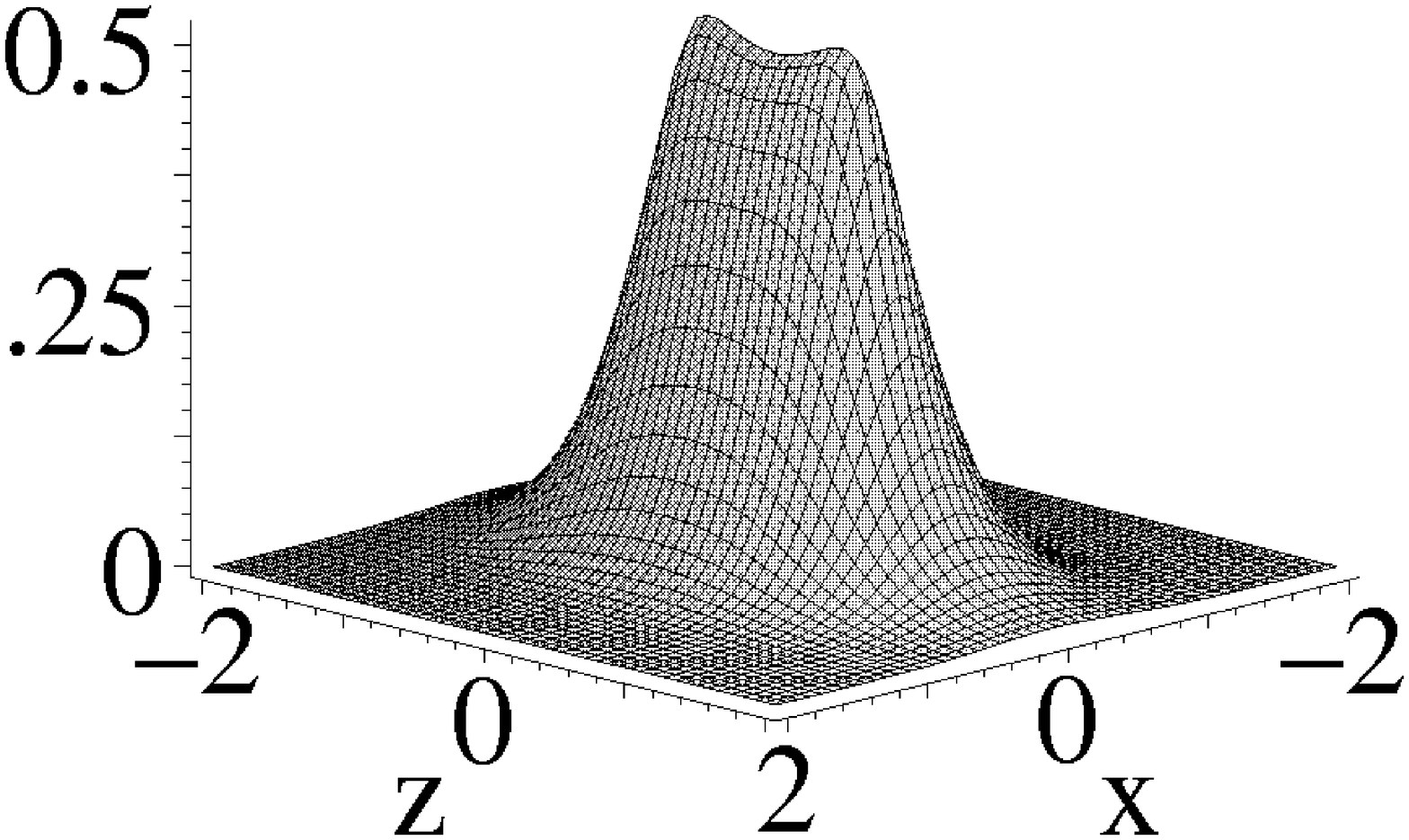}&
    \includegraphics[width=1.2in,angle=-90]{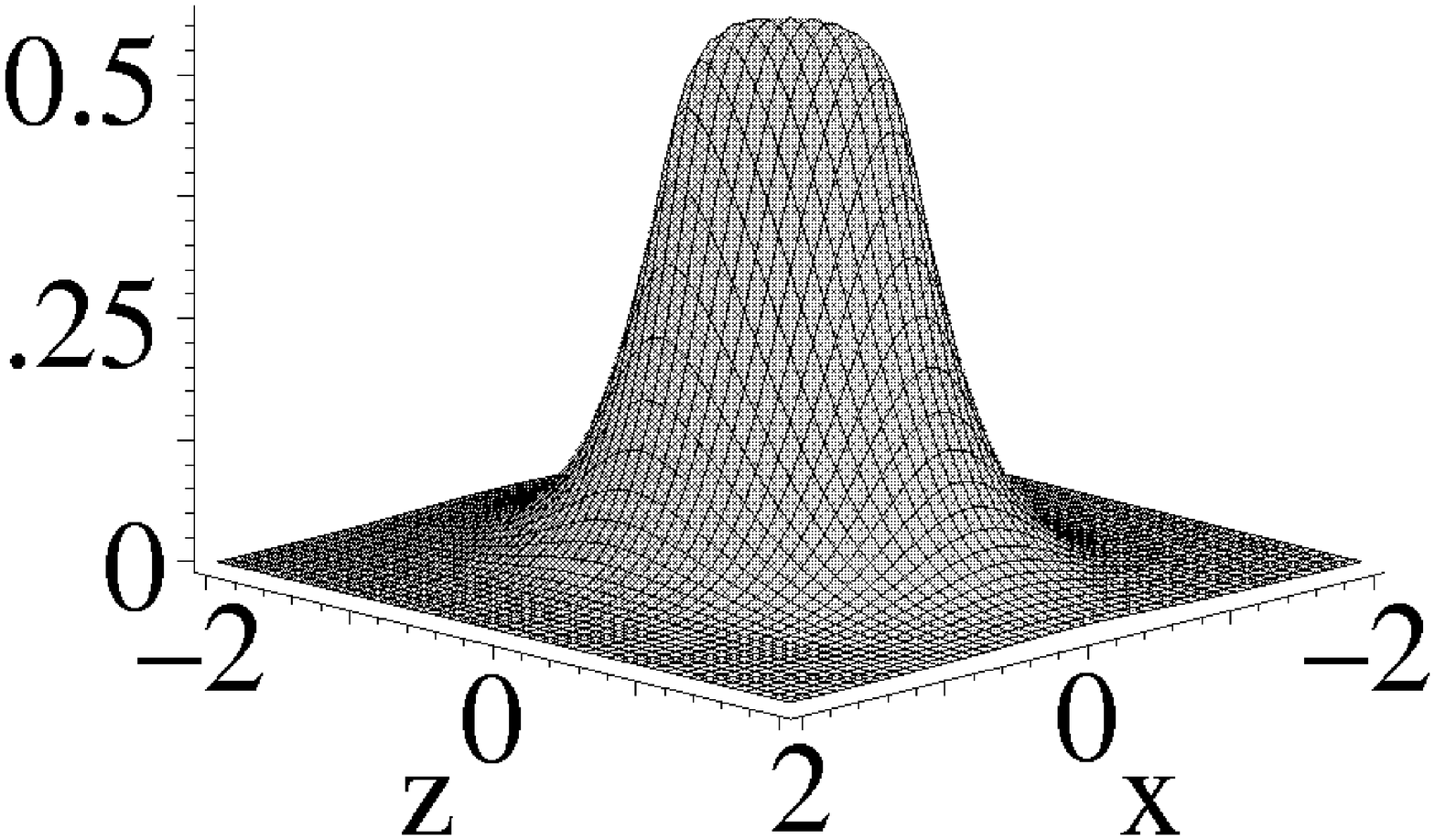} &
    \includegraphics[width=1.2in,angle=-90]{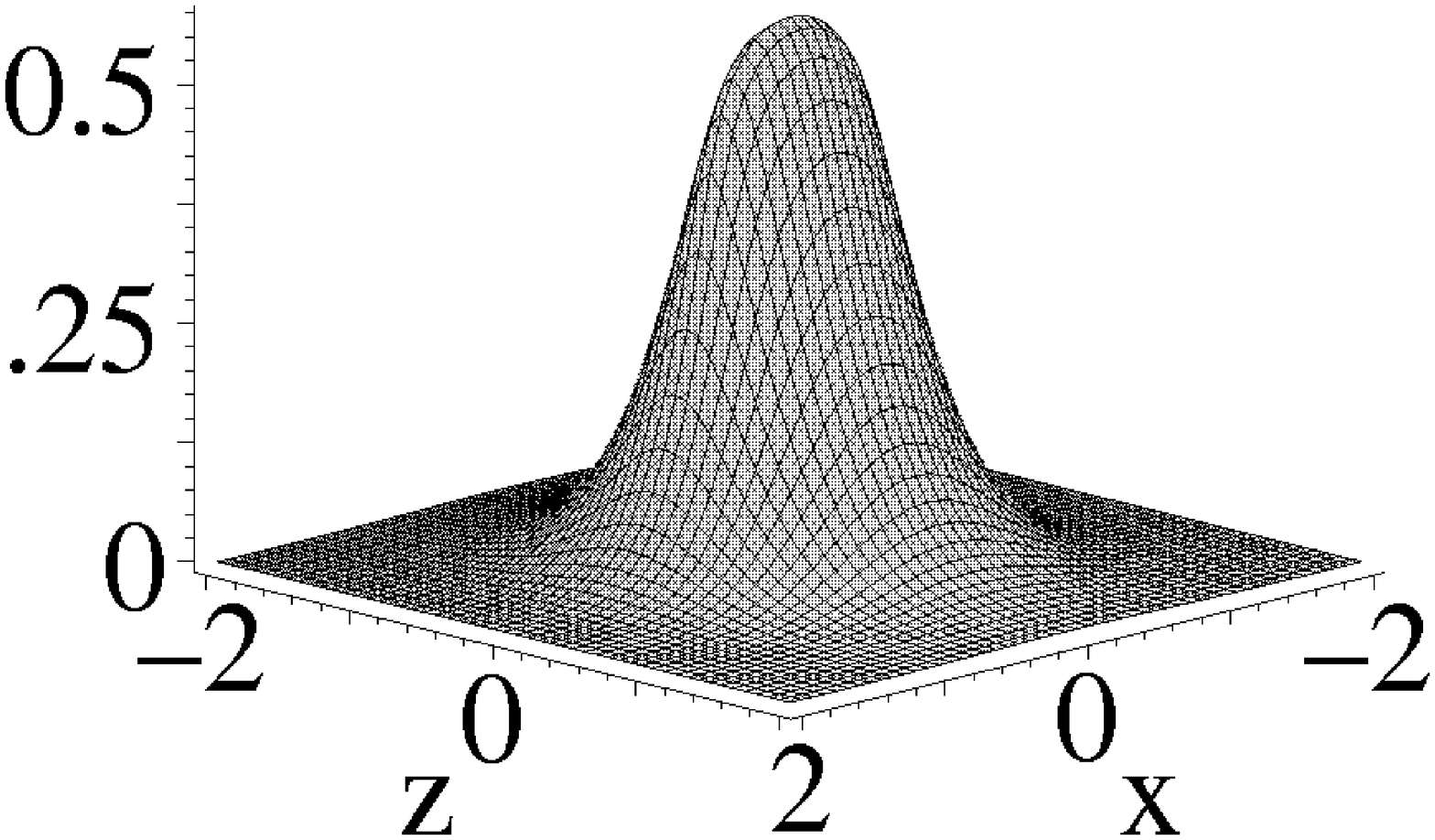} \\[5pt]
     \theta=0^{\circ} & \theta=45^{\circ} & \theta=90^{\circ} \\[15pt]
    & B=10^{11} \textrm{G} & \\
    \includegraphics[width=1.2in,angle=-90]{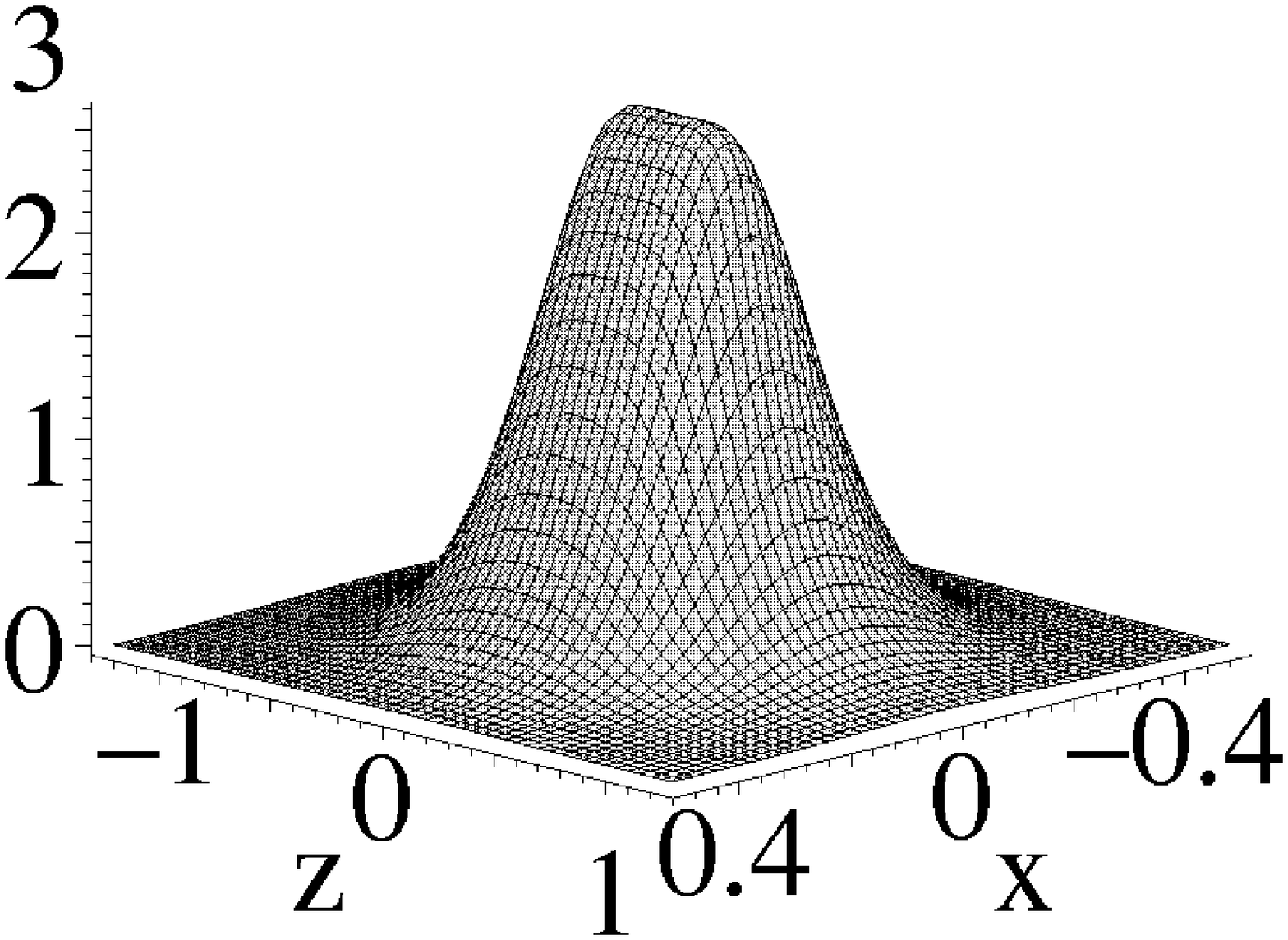} &
    \includegraphics[width=1.2in,angle=-90]{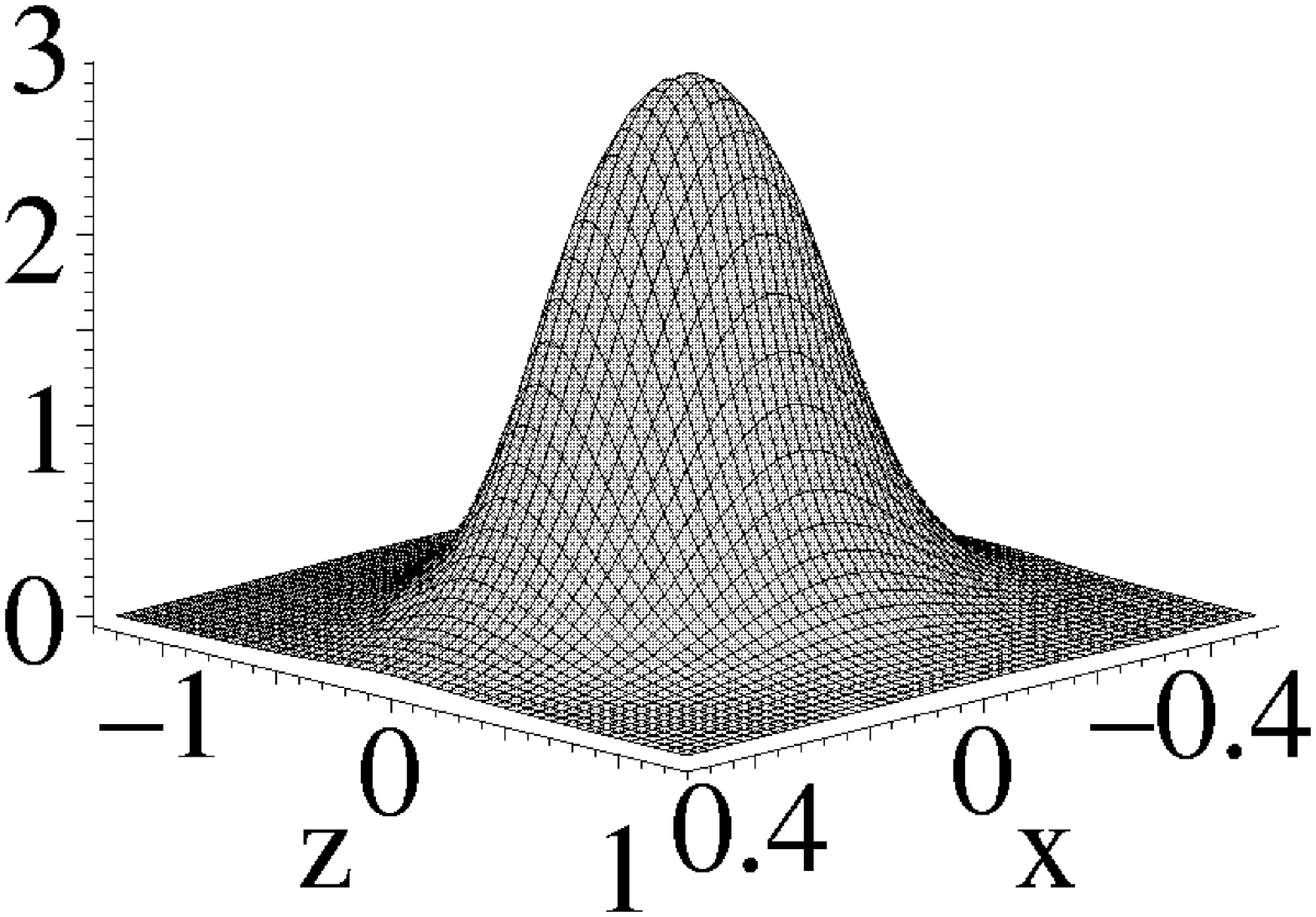}&
    \includegraphics[width=1.2in,angle=-90]{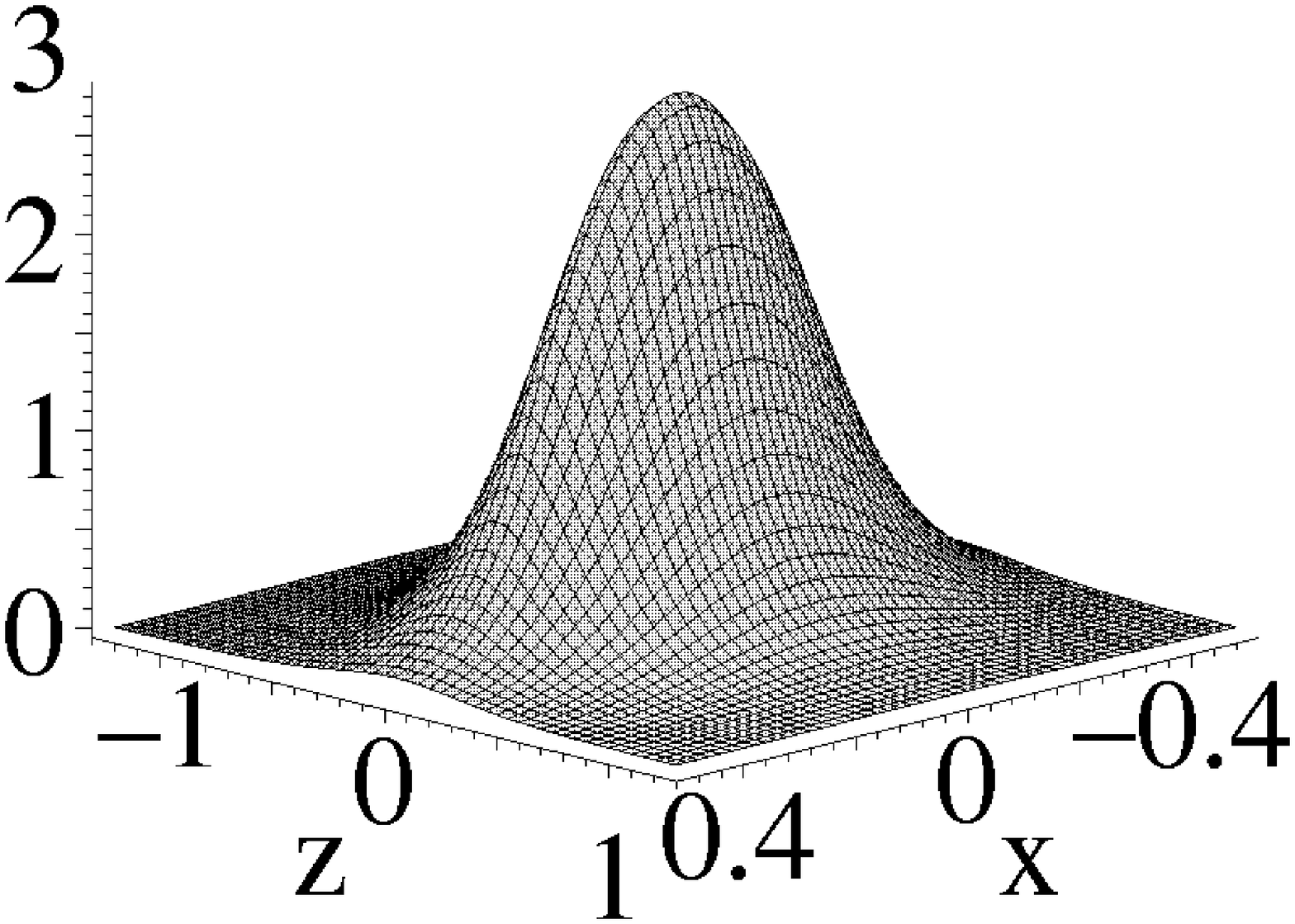} \\[5pt]
     \theta=0^{\circ} & \theta=45^{\circ} & \theta=90^{\circ} \\[15pt]
    & B=10^{12} \textrm{G} & \\
    \includegraphics[width=1.2in,angle=-90]{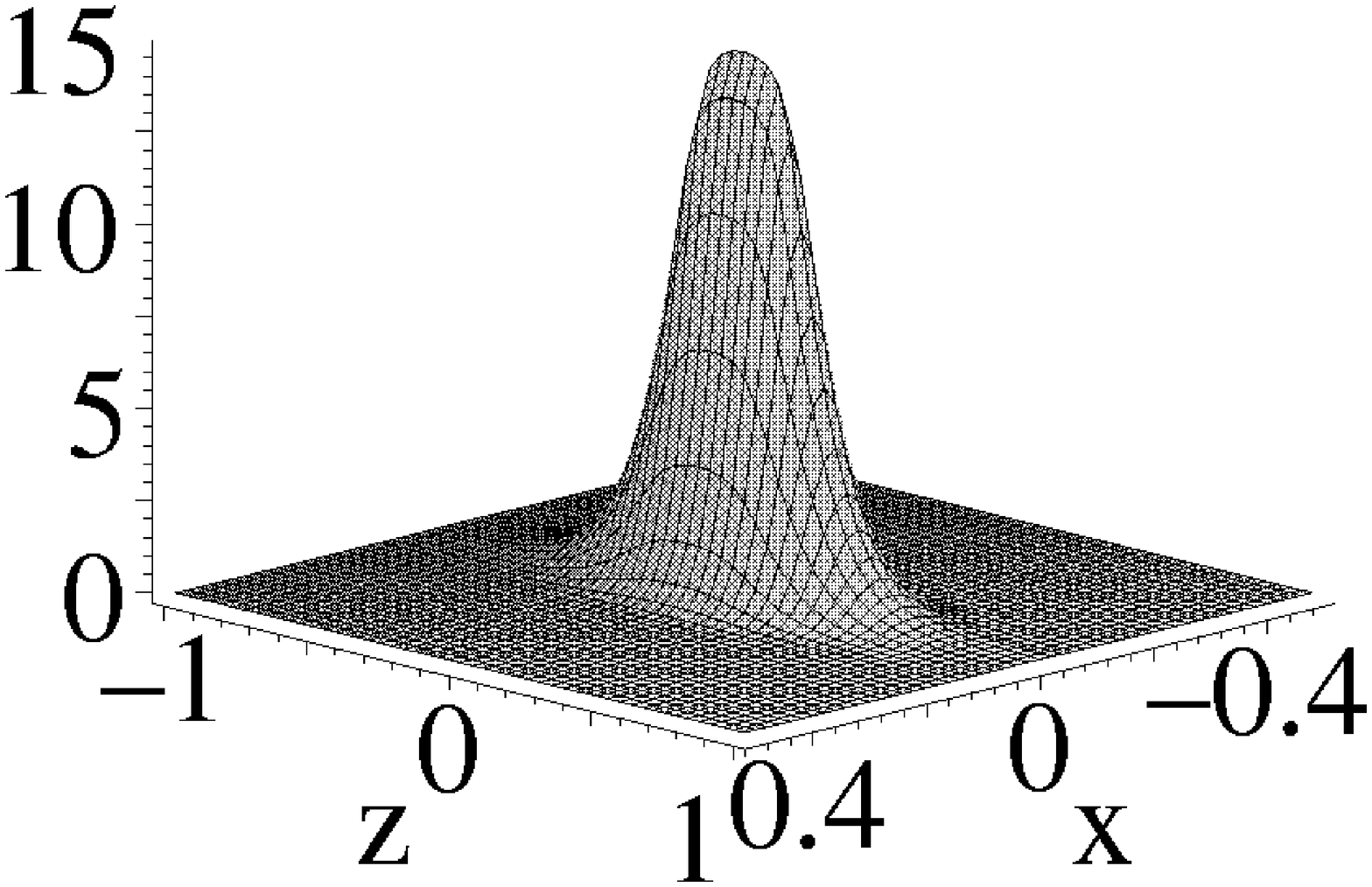} &
    \includegraphics[width=1.2in,angle=-90]{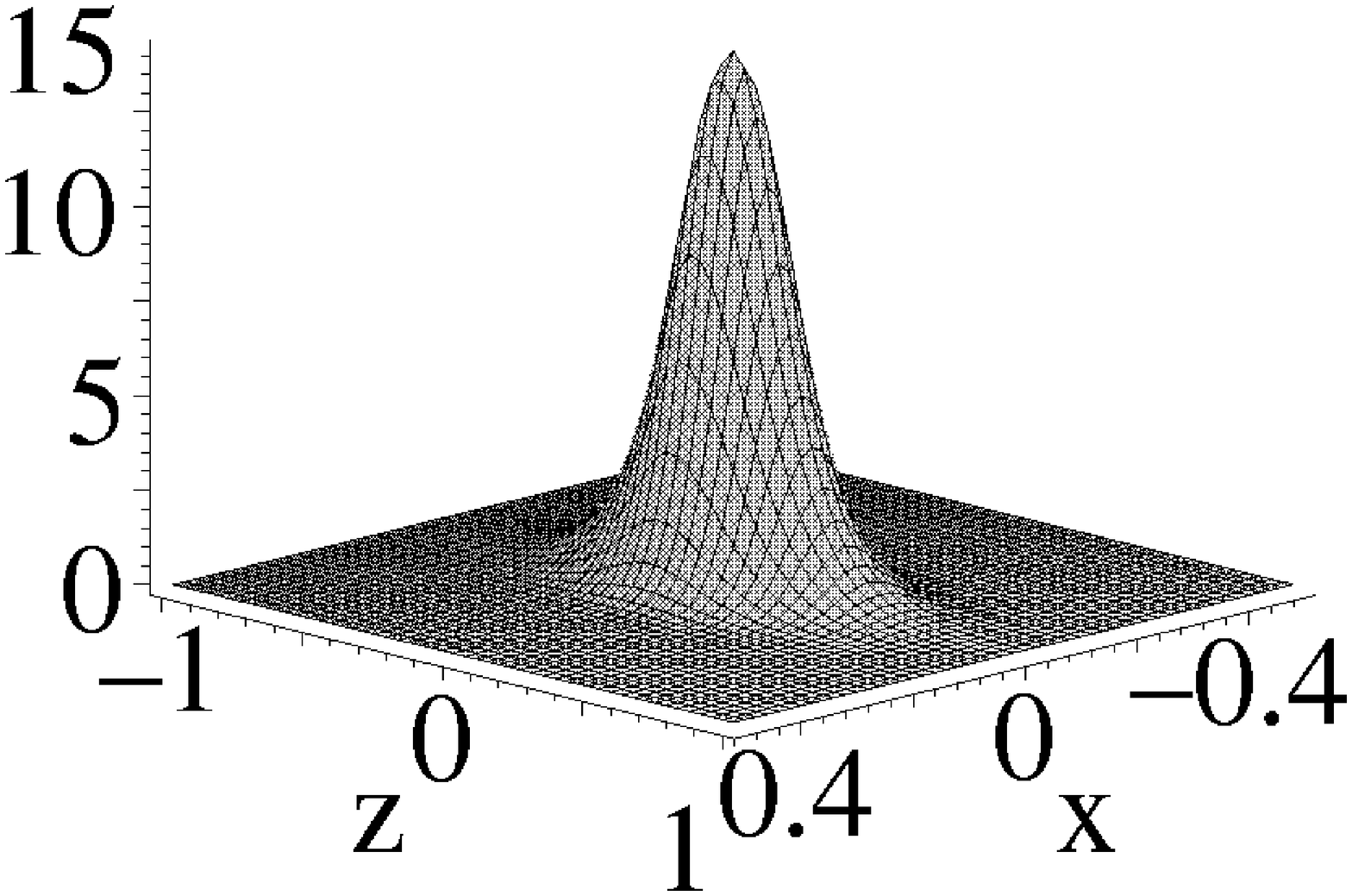} &
    \includegraphics[width=1.2in,angle=-90]{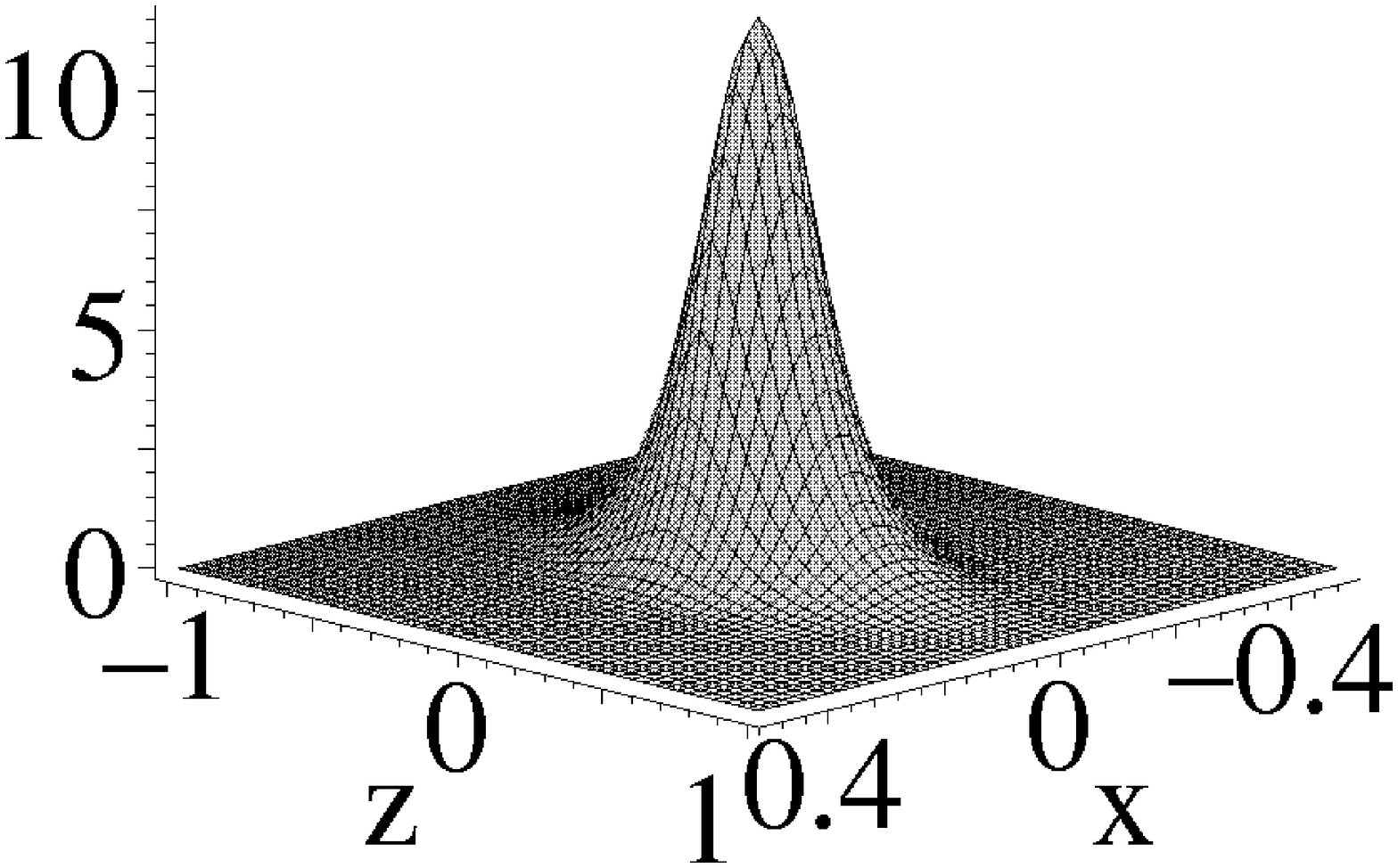} \\[5pt]
     \theta=0^{\circ} & \theta=45^{\circ} & \theta=90^{\circ} 
    \end{array}
     \]\vskip -10pt
    \caption{\label{fig:6} $H_2^+$ electronic distributions $|\psi(x,z)|^2$
      (normalized to one) for the $1_g$ state for different magnetic
      fields and inclinations.}
  \end{center}
\end{figure*}

To complete the study of the $1_g$ state we show in Fig.7 the behavior
of the variational parameters of (\ref{trial}) as a function of the
magnetic field strength for the optimal (parallel) configuration,
$\tha=0$. In general, the behavior of the parameters is rather smooth
and {\it very} slow-changing even though the magnetic field changes by
several orders of magnitude. In our opinion it reflects the level of
adequacy (or, in other words, the level of quality) of our trial
function. In practice, the parameters can be approximated by splain
method and then can be used to study any magnetic field strength other
than presented here.

\begin{figure}
\begin{center}
\[
\begin{array}{c}
\begin{picture}(215,160)(-10,0)
\put(5,20){$10^9$}
\put(40,20){$10^{10}$}
\put(83,20){$10^{11}$}
\put(125,20){$10^{12}$}
\put(165,20){$10^{13}$}
\put(85,8){$B (G)$}
\put(-10,32){$-4$}
\put(-10,62){$-3$}
\put(-10,92){$-2$}
\put(-10,122){$-1$}
\put(-10,152){$\,\,\, 0$}
\put(150,120){$A_2$}
\put(110,90){$A_1$}
\put(-10,165){\includegraphics*[width=2.3in,angle=-90]{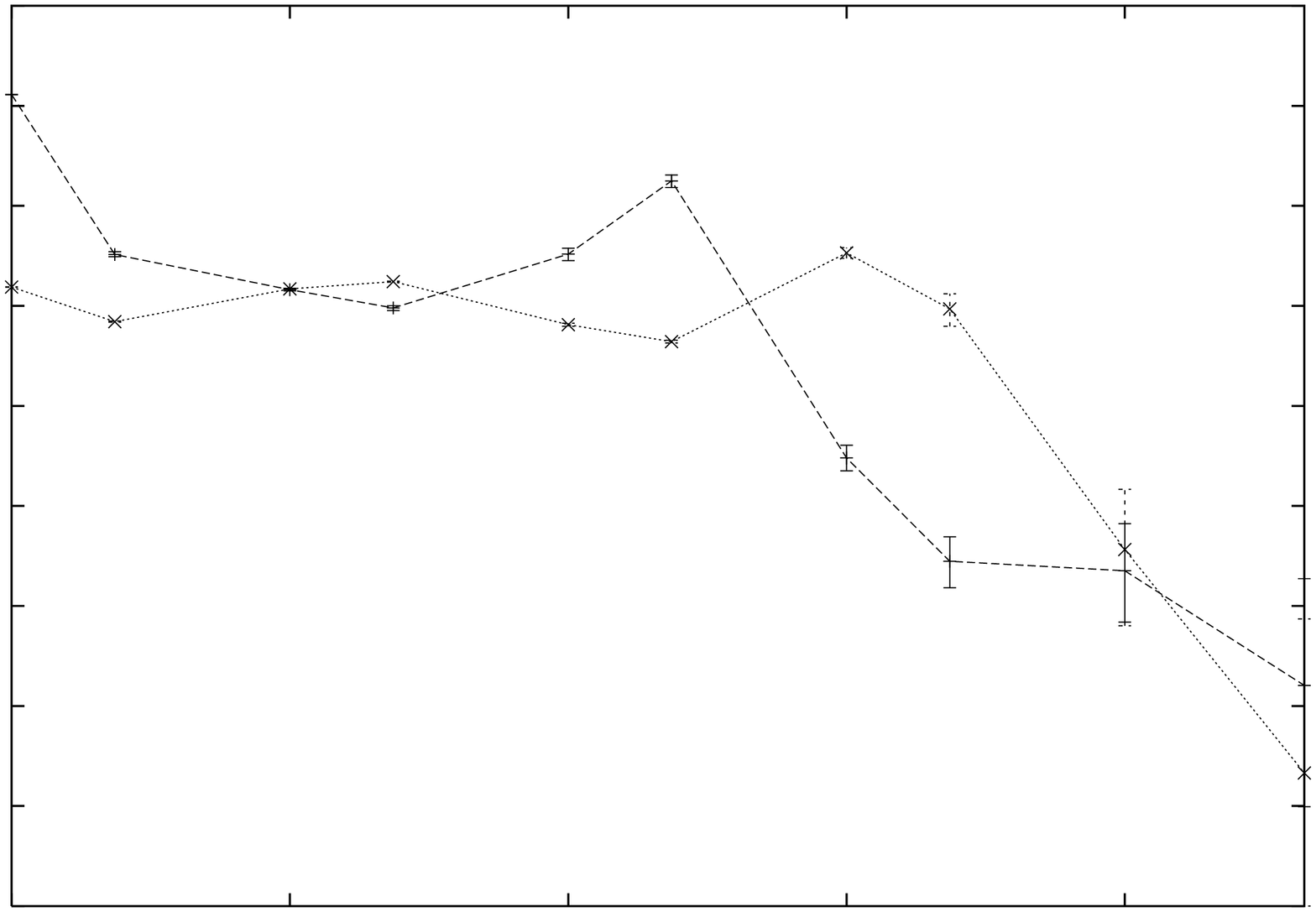}}
\end{picture}
\\[-20pt]
\begin{picture}(215,263)
\put(10,20){$10^9$}
\put(45,20){$10^{10}$}
\put(88,20){$10^{11}$}
\put(130,20){$10^{12}$}
\put(170,20){$10^{13}$}
\put(90,8){$B (G)$}
\put(5,35){$0$}
\put(5,85){$4$}
\put(5,130){$8$}
\put(4,180){$12$}
\put(4,225){$16$}
\put(-22,155){$[a.u.]^{-1}$}
\put(170,180){$\alpha_3$}
\put(160,130){$\alpha_2$}
\put(150,80){$\alpha_1$}
\put(140,42){$\alpha_4$}
\put(-22,0){\includegraphics*[width=3.35in,height=3.65in]{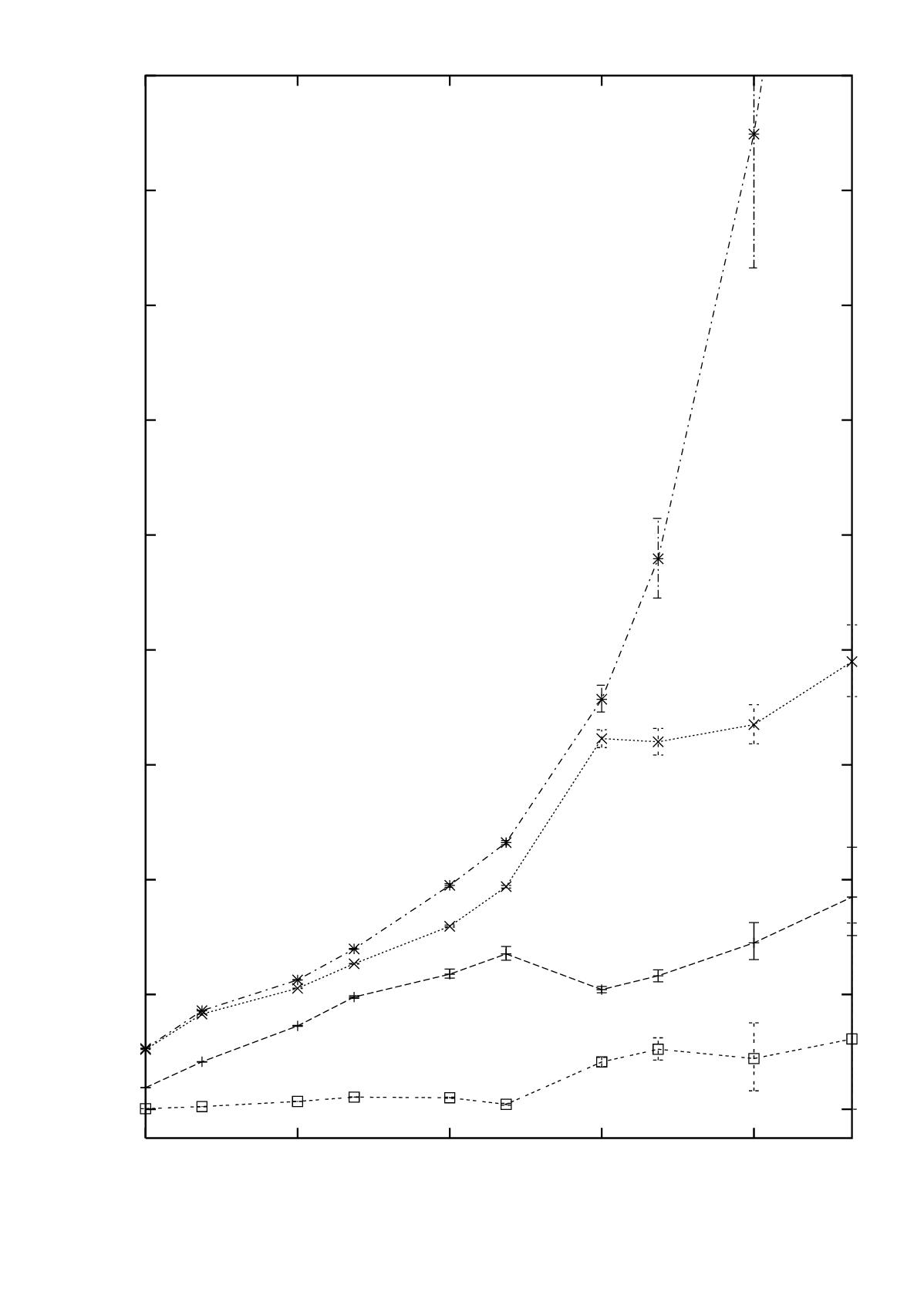}}
\end{picture}
\\[-20pt]
\begin{picture}(215,180)(-10,0)
\put(5,20){$10^9$}
\put(40,20){$10^{10}$}
\put(83,20){$10^{11}$}
\put(125,20){$10^{12}$}
\put(165,20){$10^{13}$}
\put(85,8){$B (G)$}
\put(-12,42){$0.2$}
\put(-12,65){$0.4$}
\put(-12,87){$0.6$}
\put(-12,110){$0.8$}
\put(-12,132){$1.0$}
\put(-12,152){$1.2$}
\put(-35,97){$[a.u.]^{-1}$}
\put(80,105){$\beta_{2x}$}
\put(70,136){$\beta_{3x}$}
\put(65,60){$\beta_{1x}$}
\put(-10,167){\includegraphics*[width=2.3in,angle=-90]{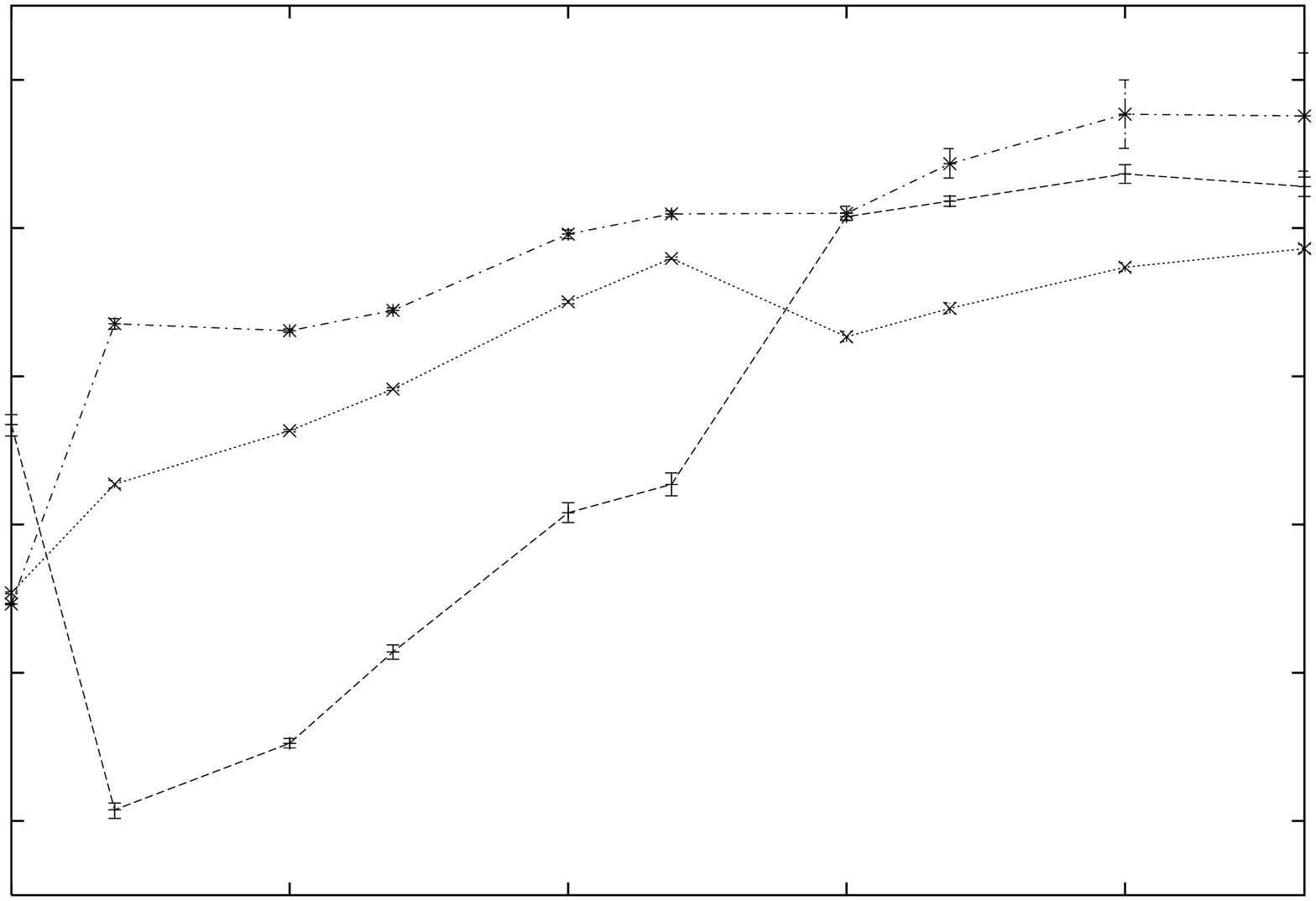}}
\end{picture} 
\end{array} 
\] \vskip -20pt
\caption{Variational parameters of the trial function (8) as function
  of the magnetic field strength $B$ for the $1_g$ state in parallel
  configuration, $\tha=0^{\circ}$. The parameter $A_3$ is fixed being
  equal to $1$ (see text). The error bars correspond to relative
  deviation in the variational energy in the region $\Delta
  E_{T}\equiv\frac{E_T}{E_{var}}\lesssim~10^{-5}$.}
\end{center} 
\end{figure}

\subsection{$1_u$ state}

In the absence of a magnetic field the $1_u$ state
\footnote{$2p\sigma_u$ in standard spectroscopic notation} is
essentially repulsive, antibonding and characterized by shallow
minimum in the total energy situated at large internuclear distance
(see, for example, \cite{Peek:1980},\cite{Schmelcher}$^{(a)}$). It is
a weakly bound state with respect to dissociation and it becomes
unbound if nuclear motion is taken into account. There are not many
studies of this state. Our major finding is that the total energy
surface of the system $(ppe)$ in the state $1_u$ exhibits a
well-pronounced minimum for $ 0 < B < 4.414 \times 10^{13}$ G and for
all inclinations.  Similar to the $1_g$ state, both total ($E_T$)
and binding ($E_b$) energies of the $1_u$ state increase as the
magnetic field grows, while the equilibrium distance shortens.
However, the accuracy of our calculations do not allow us to make a
definitive conclusion about the stability of the system with respect
to dissociation and nuclear motion effects.

Variational results for the $1_u$ state at $0^{\circ}, 45^{\circ}$ and
$90^{\circ}$ for magnetic fields $B=0 - 4.414 \times 10^{13}$ G are
shown in Tables VI-VIII.  The immediate conclusion is that
$E_T(0^{\circ})<E_T(45^{\circ})< E_T(90^{\circ})$ for all magnetic
fields. Hence, as for the $1_g$ state, the highest molecular stability
of the $1_u$ state occurs for the parallel configuration. Also, the
binding energy growth is maximal as a function of magnetic field for
the parallel configuration.  Therefore, the stability of $H_2^+$ in
the parallel configuration in the $1_u$ state even increases as the
magnetic field grows, similarly to what happens for the $1_g$ state.
These results suggest the following picture for appearance of bound
state: for small magnetic fields the minimum in the total energy
arises at very large internuclear distances \footnote{It is natural to
  assume that for $B=0$ a minimum exists at infinite (or almost
  infinite) internuclear distance.},  then, as the  magnetic field
grows, the position of the minimum moves to smaller and smaller
internuclear distances.

\begin{table*}
  \caption{ \label{Table:6} Total $E_T$,  binding $E_b$ energies  
    and equilibrium distance $R_{eq}$ for the $1_u$ state in the parallel
    configuration, $\tha=0^{\circ}$.  ${}^{\dagger}$This value is
    taken from \cite{Lopez:1997} }
\begin{ruledtabular}
    \begin{tabular}{lcccl}
\( B \) &  \( E_T \) (Ry)& \( E_{b} \) (Ry) & \( R_{eq} \) (a.u.)& \\
\hline
$B=0$         & -1.00010   & 1.00010  & 12.746
              &Present${}^{\dagger}$\\
              & -1.00012   & 1.00012   & 12.55   &Peek et al\\
 \( 10^{9} \) G
           & -0.92103   & 1.34656 & 11.19  & Present \\
           & -0.917134  &  ---    & 10.55  & Peek et al
           \cite{Peek:1980}\\[5pt]
 \( 1 \) a.u.
              &  -0.66271  & 1.66271   & 9.732  & Present\\
              &  -0.66     & 1.66      & 9.6    & Kappes et al
              \cite{Schmelcher}\\[5pt]
\( 10^{10} \) G
               &  1.63989  & 2.61500  & 7.180  & Present \\
               &  2.1294   &  ---     & 4.18   & Peek et al
               \cite{Peek:1980}\\[5pt]
 \( 10 \) a.u.
               &  6.52362  & 3.47638   & 6.336   & Present\\[5pt]
 \( 10^{11} \)G
               & 36.83671  & 5.71649   & 4.629  & Present \\[5pt]
 \( 100 \) a.u.
               & 92.42566  & 7.57434   & 3.976  & Present\\[5pt]
 \( 10^{12} \)G
               & 413.6175  & 11.9144   & 3.209  & Present \\[5pt]
 \( 1000 \) a.u.
               & 984.6852  & 15.3148   & 2.862  & Present\\[5pt]
 \( 10^{13} \)G
               & 4232.554  & 22.7648   & 2.360  & Present \\[5pt]
\( 4.414\times 10^{13} \) G
               & 18750.07  & 32.9104   & 1.794  & Present \\[5pt]
\end{tabular}
\end{ruledtabular}
\end{table*}

\begin{table}
  \caption{\label{Table:7} Total $E_T$, binding $E_b$ energies  
  and equilibrium distance $R_{eq}$ for the $1_u$ state in the 
   configuration $\tha=45^{\circ}$.
   Optimal value for the gauge  parameter $\xi$ is shown (see text).}
\begin{ruledtabular}
    \begin{tabular}{lcccc}
\( B \)&  \( E_T \) (Ry)& \( E_{b} \) (Ry)& \( R_{eq} \) (a.u.)&\( \xi  \)\\
\hline
 \(10^{9}\)G
             & -0.870391 & 1.295923  & 8.053   & 0.9308  \\[5pt]
 \( 1 \) a.u.
             & -0.509041 & 1.509041  & 6.587   & 0.9406  \\[5pt]
 \(10^{10}\) G
             &  2.267998 & 1.987321  & 4.812   & 0.9671  \\[5pt]
 \( 10 \) a.u.
             &  7.692812 & 2.307188  & 4.196   & 0.9808  \\[5pt]
  \(10^{11}\) G
             & 39.71061  & 2.84258   & 3.538   & 0.9935  \\[5pt]
  \( 100 \) a.u.
             & 96.88464  & 3.11536   & 3.278   & 0.9968  \\[5pt]
  \(10^{12}\) G
             & 422.0074  & 3.5245    & 3.020   & 0.9991  \\[5pt]
  \( 1000 \) a.u.
             & 996.3044  & 3.6956    & 2.894   & 0.9996  \\[5pt]
  \(10^{13}\) G
             & 4251.409  & 3.9103    & 2.790   & 0.9999  \\[5pt]
  \( 4.414\times 10^{13} \) G
             & 18778.95  & 4.0330    & 2.746   & 0.9999  \\[5pt]
\end{tabular}
\end{ruledtabular}
\end{table}

\begin{table*}
  \caption{\label{Table:8} Total $E_T$,
  binding $E_b$ energies and equilibrium distance
  $R_{eq}$  for the $1_u$ state at $\tha=90^{\circ}$.  Optimal value
  for the gauge   parameter $\xi$ is shown (see text). }
\begin{ruledtabular}
    \begin{tabular}{lccccl} 
\( B \)&  \( E_T \) (Ry)& \( E_{b} \) (Ry)& \( R_{eq} \) (a.u.)&\( \xi  \)& \\
\hline 
\(10^{9}\) G
 & -0.867234  & 1.292766  & 8.784  & 0.9692  & Present  \\[5pt]
\(1\) a.u.
 & -0.49963   & 1.49963   & 7.264  & 0.9737  & Present  \\
 & -0.65998   & 1.65998   & 5.45   &         & Kappes et al
 \cite{Schmelcher}(b)  \\[5pt]
\(10^{10}\) G
 &  2.29365   & 1.96167   & 5.517  & 0.9866  & Present  \\[5pt]
\(10\) a.u.
 &  7.72998   & 2.27002   & 4.872  & 0.9923  & Present  \\[5pt]
\(10^{11}\) G
 & 39.76500   & 2.78819   & 4.154  & 0.9975  & Present  \\[5pt]
\(100\) a.u.
 & 96.93497   & 3.06503   & 3.875  & 0.9988  & Present  \\[5pt]
\(10^{12}\) G
 & 422.0834   & 3.44848   & 3.594   & 0.9997  & Present  \\[5pt]
\(1000\) a.u.
 & 996.3807   & 3.61935   & 3.460   & 0.9998  & Present  \\[5pt]
\(10^{13}\) G
 & 4251.497   & 3.82238  & 3.340   & 0.9999  & Present  \\[5pt]
\( 4.414\times 10^{13} \) G
 &  18779.04  & 3.9409   & 3.306   & 0.9999  & Present  \\[5pt]
\end{tabular}
\end{ruledtabular}
\end{table*}

Our results for $B>0$ and $\tha=0^{\circ}$ give the lowest total
energies compared to other calculations. They are in good agreement
with those by Kappes--Schmelcher \cite{Schmelcher}$^{(a)}$ as well as
by Peek--Katriel \cite{Peek:1980} for $B=0, 10^9$ G, although for
$B=10^{10}$ G a certain disagreement is observed (see Table VI).
However, for $\tha=90^{\circ}$ our results are in striking contrast
with those by Wille \cite{Wille:1988}, where even the optimal
configuration is attached to $\tha=90^{\circ}$, contrary to our
conclusion. For instance, at $B=10^{10}$ G in \cite{Wille:1988} the
values $E_b=2.593$ Ry and $R_{eq}=2.284$ a.u. are given, while our
results are $E_b=1.9617$ Ry and $R_{eq}=5.517$ a.u., respectively (see
Table VIII). Similar, but less drastic disagreement, is the observed
with the results in \cite{Schmelcher}$^{(b)}$.  We can guess this
disagreement is due to the shallow nature of the minimum, but a real
explanation of this effect is missing.

The analysis of Tables VI-VIII shows that for $\tha>0^{\circ}$ and
fixed magnetic field the total energy of $H_2^+$ in the $1_u$ state is
always larger than the total energy of the hydrogen atom
\cite{Kravchenko}.  It means that the $H_2^+$-ion in the $1_u$ state
is unstable towards dissociation to $H + p$.  For $\tha\sim 0^{\circ}$
the present total energies of the $H_2^+$ ion and the most accurate
results for the hydrogen atom \cite{Kravchenko} are comparable within
the order of magnitude $10^{-4} - 10^{-5}$. At the same time we
estimate that the accuracy of our calculations is of the same order of
magnitude. It prevents us from making a conclusion about the stability
of $H_2^+$ in the $1_u$ state with respect to dissociation. The only
conclusion can be drawn is that the minimum is very shallow.

The $1_u$ state is much more extended than the $1_g$ state: for fixed
magnetic field the equilibrium distance of the $1_g$ state is much
smaller than that for the $1_u$ state. This picture remains the same
for any inclination. It is quite striking to see the much lower rate of
decrease of $R_{eq}$ in the range $B=0 - 4.414 \times 10^{13}$ G: for
the state $1_u$ it falls  $\sim 3$ times compared to the $1_g$
state, where it falls  $\sim 20$ times.

The behavior of the equilibrium distance $R_{eq}$ of the $1_u$ state
as a function of inclination is quite non-trivial (see Tables
VI-VIII). As in the $1_g$ state, the H$_2^+$-ion in the $1_u$ state
for $B \lesssim 10^{12}$ G is most extended in the parallel
configuration, while being most compact at $\tha \simeq 45^{\circ}$.
For $B \sim 10^{12}$ G the equilibrium distance is almost independent
of the inclination, while for $B > 10^{12}$ G the most compact
configuration corresponds to $\tha=0^{\circ}$, in contrast to the
$1_g$ state.

\section{Conclusion}

We carried out accurate non-relativistic calculations in
the Bohr-Oppenheimer approximation for the lowest states of the H$_2^+$
molecular ion of even parity, $1_g$, and odd parity, $1_u$, for
different orientations of the magnetic field with respect to the
molecular axis. We studied constant magnetic fields ranging up to $B =
4.414 \times 10^{13}$ G, where a non-relativistic consideration is valid.

For all studied magnetic fields and orientations a well-pronounced
minimum in the total energy surface for both $1_g$ and $1_u$ states
was found. This makes manifest the existence of $H_2^+$ in both states
for magnetic fields $B = 0- 4.414 \times 10^{13}$ G. The smallest
total energy was always found to correspond to the parallel
configuration $\tha=0^{\circ}$. The total energy increased while the
binding energy decreased steadily as the inclination angle grew from
$0^{\circ}$ to $90^{\circ}$ for both states. The rate of the total
energy increase as well as the binding energy decrease was seen to be
always maximal for the parallel configuration for both states.

However, the equilibrium distance exhibited quite non-trivial behavior
as a function of the orientation angle $\tha$.  In the case of the
$1_g$ state, the shortest equilibrium distance always corresponded to
the perpendicular configuration for magnetic fields $B \lesssim
10^{12}$ G, whereas for $B \gtrsim 10^{12}$ it occurred for some angle
$\tha<90^{\circ}$. On the contrary for the $1_u$ state the shortest
equilibrium distance always corresponded to orientations $\tha <
90^{\circ}$ for all magnetic fields considered. In particular, for $B
> 10^{12}$ G, it begins to correspond to the parallel configuration.
As for the largest equilibrium distances -- they were found to always
correspond to the parallel configuration for both the $1_g$ state for
all studied magnetic fields and for the $1_u$ state for $B \lesssim
10^{12}$.  However, for larger magnetic fields $B \gtrsim 10^{12}$ the
largest equilibrium distance for the $1_u$ state was seen to
correspond to the perpendicular configuration (!).

Confirming the previous qualitative observations made by Larsen
\cite{Larsen} and Khersonskij \cite{Kher} for the $1_g$ state we
demonstrated that the H$_2^+$ ion in the lowest energy state can
dissociate to $H + p$ for a certain range of orientations starting
from magnetic fields B$\gtrsim 1.8 \times 10^{11}$ G. As the magnetic
field increases the region where dissociation is allowed was seen to
steadily broaden, reaching $25^{\circ}\lesssim \tha \leqslant
90^{\circ}$ for $B=4.414\times 10^{13}$ G.

The electronic distributions were found to be qualitatively different
for weak and large magnetic fields. In the domain $B<10^{10}$ G the
electronic distribution for any inclination peaks near the position of
the protons. On the contrary for $B>10^{11}$ G the electronic
distribution is always peaked near the midpoint between the protons
for any inclination. It implies physically different structure of the
ground state - for weak fields the ground state can be modeled as a
combination of hydrogen atom and proton while for strong fields such
modeling is irrelevant.

Combining all the above-mentioned observations we conclude that for
magnetic fields of the order of magnitude $B \sim 10^{11}$ G some
qualitative changes in the behavior of the $H_2^+$ take place. The
behavior of the variational parameters also favors this conclusion. It
looks like as appearance of a new scale in the problem. It might be
interpreted as a signal of a transition to the domain of developed
quantum chaos \cite{chaos}.

\begin{acknowledgments}
  A.V.T. and J.C.L.V. thank Labo.Phys.Theo., Universit\'e Paris Sud
  and ITAMP, Harvard University for their kind hospitality extended to
  them, where a part of the present work was done. We want to express
  our deep gratitude to anonymous referee for constructive criticism
  which helped to improve presentation of the article. The authors
  thank C. Stephens for careful reading of the manuscript. This work
  was supported in part by DGAPA Grant \# IN120199 (Mexico).  A.V.T.
  thanks NSF grant for ITAMP at Harvard University and CONACyT grant
  for a partial support. A.F.R. would like to acknowledge a partial
  provision of funds extended by CONACyT (Mexico) through project
  32213-E.
\end{acknowledgments}

\end{document}